\documentclass{aa}  
\usepackage{graphicx}
\usepackage{txfonts}
\newcommand{\ros}{\textit{ROSAT}}
\newcommand{\chan}{\textit{Chandra}}
\newcommand{\xmm}{\textit{XMM-Newton}}
\newcommand{\eROS}{\textit{eROSITA}}

\newcommand{\eso}{ESO-VLT}

\newcommand{\nh}{N_{\rm H}}
\def \etacar{\object{Eta~Carinae}}
\def \msev{M7}

\def \magze{\object{RX~J0806.4-4123}}

\def \jtenfull{\object{2XMM~J104608.7-594306}}
\def \jten{J1046}

\def \calvfull{\object{1RXS~J141256.0+792204}}
\def \runaway{\object{MJ~218}}
\begin{document} 

%
%
\title{New \xmm\ observation of the thermally emitting isolated neutron star \jtenfull 
       \thanks{Based on observations obtained with \xmm, an ESA science mission with instruments and contributions directly funded by ESA Member States and NASA (Target \jtenfull, \textsf{\small obsids}~0650840101, 0691970101).}}
\author{A.~M.~Pires\inst{1}
    \and C.~Motch\inst{2}
    \and R.~Turolla\inst{3,4}
    \and S.~B.~Popov\inst{5}
    \and A.~D.~Schwope\inst{1}
    \and A.~Treves\inst{6}}
\offprints{A. M. Pires}
\institute{Leibniz-Institut f\"ur Astrophysik Potsdam (AIP), An der Sternwarte 16, 14482 Potsdam, Germany, 
    \email{apires@aip.de} 
    \and
    Observatoire Astronomique de Strasbourg, Universit\'e de Strasbourg, CNRS, UMR 7550, 11 rue de l'Universit\'e, F-67000 Strasbourg, France
    \and
    Universit\'a di Padova, Dipartimento di Fisica e Astronomia, via Marzolo 8, 35131 Padova, Italy
    \and 
    Mullard Space Science Laboratory, University College London, Holmbury St. Mary, Dorking, Surrey, RH5 6NT, UK
    \and
    Sternberg Astronomical Institute, Lomonosov Moscow State University, Universitetskii pr. 13, 119991 Moscow, Russia
    \and
    Universit\'a dell'Insubria, Dipartimento di Fisica e Matematica, Via Valleggio 11, 22100 Como, Italy}   
\date{Received ...; accepted ...}
\titlerunning{The isolated neutron star in the Carina nebula revisited}
\authorrunning{A.~M.~Pires et al.}
\abstract
{The isolated neutron star (INS) \jtenfull\ is one of the only two to be discovered through their thermal emission since  the \ros\ era. Possibly a remnant of a former generation of massive stars in the Carina nebula, the exact nature of the source is unclear, and it might be unique amongst the several classes of galactic INSs.}
{In a first dedicated \xmm\ observation of the source, we found intriguing evidence of a very fast spin period of $P\sim18.6$\,ms at the $4\sigma$ confidence level. Moreover, spectral features in absorption have also been identified. We re-observed \jtenfull\ with \xmm\ to better characterise the spectral energy distribution of the source, confirm the candidate spin period, and possibly constrain the pulsar spin-down.}
{We used the two \xmm\ observations of \jtenfull\ to perform detailed timing and spectral X-ray analysis. Both the spin-down rate and the energy of the spectral features provide estimates on the neutron star magnetic field, which are crucial for investigating the evolutionary state of the neutron star.}
{Statistically acceptable spectral fits and meaningful physical parameters for the source are only obtained when the residuals at energies $0.55$\,keV and $1.35$\,keV are taken into account by the spectral modelling. While the former can result from the inhomogeneous temperature distribution on the surface of the neutron star or can be related to a local overabundance of oxygen in the Carina nebula, the one at 1.35\,keV is only satisfactorily accounted for by invoking a line in absorption. In this case, the best-fit neutron star atmosphere models constrain the hydrogen column density, the effective temperature, and the luminosity of the source within $\nh=(2.5-3.3)\times10^{21}$\,cm$^{-2}$, $T_{\rm eff}=(6-10)\times10^5$\,K, and $L_{\rm X}=(1.1-7.4)\times10^{32}$\,erg\,s$^{-1}$. The implied distance is consistent with a location in (or in front of) the Carina nebula, and radiation radii are compatible with emission originating on most of the surface. Non-thermal X-ray emission is ruled out at levels above $0.5\%$ ($3\sigma$) of the source luminosity. 
Unfortunately, the second \xmm\ observation proved inconclusive in terms of confirming (discarding) the fast candidate spin, providing an upper limit on the pulsed fraction of the source that is very close to the limiting sensitivity for detecting the modulation found previously.} 
{In the absence of an unambiguous period determination and an estimate of the magnetic field, the nature of the source remains open to interpretation. Its likely association with the Carina cluster and its overall spectral properties (only shared by a handful of other peculiar INSs) disfavour a standard evolutionary path or one in which the source was previously recycled by accretion in a binary system. The star \jtenfull\ may be similar to Calvera (\calvfull), a neutron star for which the scenario of an evolved anti-magnetar has been discussed. A better age estimate and deeper radio and $\gamma$-ray limits are required to further constrain the evolutionary state of the neutron star.}

\maketitle
\section{Introduction\label{sec_intro}}
In the standard scenario of pulsar evolution, a neutron star is born rotating fast. As a consequence of rotational energy losses, it is expected to evolve towards longer periods until electron-positron pairs can no longer be produced; the radio emission ceases and the pulsar crosses the so-called death line. The radio emission is, however, energetically unimportant relative to the power emitted at high energies. While young, these objects are strong sources of X-ray and $\gamma$-ray emission with radiative processes that include non-thermal emission from charged particles that were accelerated in the pulsar magnetosphere and thermal emission from the hot surface. However, after a maximum of $\sim10^7$\,yr, pulsars have probably exhausted their internal source of particle creation and acceleration and cooled down to temperatures undetectable in X-rays.

This relatively simple picture has been challenged by the discovery of peculiar groups of isolated neutron stars (INSs) that are radio quiet or transient. Their (mostly unpredicted) properties significantly differ from those displayed by standard pulsars \citep[see][for an overview and references]{kas10,har13}. Among them are magnetars (anomalous X-ray pulsars (AXPs) and soft gamma repeaters (SGRs)), the central compact objects in supernova remnants (CCOs), the rotating radio transients (RRATs), and the \ros-discovered thermally-emitting INSs, also known as the Magnificent Seven (\msev). 

In particular, the \msev\ constitute a nearby ($d\lesssim1$\,kpc) group of middle-aged ($0.1-1$\,Myr) neutron stars, observable through their bright and purely thermal X-ray emission \citep[see, for example,][for reviews]{hab07,kap08a,tur09}. Amounting to about half of all young INSs known in the solar neighbourhood \citep{pop03}, they could be as numerous as radio pulsars, with consequences for the total number of neutron stars populating the Milky Way \citep[see][]{kea08}.
Timing studies in X-rays \citep[see][and references therein]{kap11b} have shown that the \msev\ rotate more slowly ($P\sim3-11$\,s) and have higher magnetic field intensities, $B_{\rm dip}\sim(1-3)\times10^{13}$\,G, than the bulk of the (rotation-powered) radio pulsar population. Also at variance with pulsars detected at high energies, the X-ray luminosity of the \msev\ is in excess of the spin-down power. To a certain extent, such properties resemble those of the young and energetic magnetars \citep[see,  for example,  ][for a review]{mer08}. On the other hand, the \msev\ experience less dramatic spin-down and do not show the complex phenomenology usually observed in magnetars (e.g.~emission of flares, bursts, quasi-periodic oscillations, spectral variability, timing noise, etc). 

Another group consisting of radio-quiet, thermally emitting INSs, is that of CCOs \citep[see, for example, ][for reviews]{luc08,got10a,hal10}. CCOs were first observed as point-like, radio-quiet X-ray sources, located near the geometrical centre of supernova remnants. Of the dozen objects known at present, including other candidates, at least three fit in the anti-magnetar scenario, in which these sources are young (aged $10^3$\,yr to $10^4$\,yr) and weakly magnetised ($B_{\rm dip}=10^{10}-10^{11}$\,G) neutron stars that are experiencing very low spin-down \citep[][and references therein]{got13a}. 

If the weak dipolar field of anti-magnetars remains constant, they are confined to a region of the $P-\dot{P}$ diagram of INSs that is devoid of other pulsars, between those occupied by the normal and millisecond (recycled) part of the population. The fact that no old (orphaned, or without a supernova remnant) CCO is recognised in this region of the $P-\dot{P}$ diagram in either radio or X-ray surveys (\citealt{hal10,got13b,luo15}; see also \citealt{bog14b}, who searched for CCO descendants among supposedly old radio pulsars in supernova remnants) may favour the alternative scenario of field burial by hypercritical accretion \citep[e.g.][]{che89,gep99,vig12}. In this scenario, the original neutron star magnetic field is submerged by fallback accretion of supernova ejecta, remaining hidden for several $\sim10^3$\,yr to $10^5$\,yr. As a result, the neutron star shows, in its early evolution, the typical anti-magnetar behaviour. As the field re-emerges, the pulsar quickly joins the rest of the population, moving upwards in the $P-\dot{P}$ diagram; its subsequent evolution would then follow according to the original field intensity. Residual thermal luminosity in the first $\sim10^4$\,yr, as well as enhanced cooling after $\sim10^5$\,yr, compared to its peers of similar characteristic ages, are expected to result from the accreted light-element envelope \citep[e.g.][]{yak04}. 

For several years, considerable efforts have been made to discover new thermally emitting INSs \citep[e.g.][]{chi05,tur10,agu11}. The INS \jtenfull\ (hereafter \jten), is one of only two to be discovered through their thermal emission and lack of obvious counterparts since the \ros\ era \citep{pir09b}. Possibly a remnant of a former generation of massive stars in the Carina nebula \citep{tow11b}, its exact nature is not clear, and it may be unique. The other source, \calvfull, also known as Calvera \citep{rut08}, is a relatively bright \ros\ source with a short spin period of $P\sim59$\,ms \citep{zan11}. Calvera is the only INS to date to be recognised as an evolved anti-magnetar, with a magnetic field of $B_{\rm dip}=4.4\times10^{11}$\,G that is, possibly, regaining its original strength \citep{hal13}. 

Thanks to its proximity (angular distance of $8.5'$) to the well-studied \etacar\ system, \jten\ was serendipitously detected on many occasions by the \xmm\ Observatory \citep{jan01}. 
The analysis of the serendipitous data shows no significant long-term variability in either the X-ray flux or in the overall spectral properties of the source \citep{pir09a}. Follow-up observations in the optical with the European Southern Observatory Very Large Telescope (\eso) set a deep limit on the brightness of the optical counterpart, $m_{\rm V}>27$ ($2\sigma$), and a very high X-ray-to-optical flux ratio of $\log(F_{\rm X}/F_{\rm V})>3.8$. At present, no radio or $\gamma$-ray counterparts are known. 

As revealed by a first dedicated \xmm\ observation, performed on 2010 December 6 (AO9, as reference for the text), the spectral energy distribution of \jten\ is soft and purely thermal with significant deviations from a blackbody continuum of $kT\sim135$\,eV \citep{pir12}.
While these properties are reminiscent of those of the middle-aged \msev, the \xmm\ observation also revealed intriguing evidence of a very fast rotation, at $P\sim18.6$\,ms. Such a rapid spin is difficult to reconcile with the spectral properties of the source. 
\begin{table}[t]
\caption{Instrumental configuration and duration of the EPIC scientific exposures of the \xmm\ AO11 observation of \jtenfull
\label{tab_exposureinfoEPIC}}
\centering
\begin{tabular}{l c c c}
\hline\hline
Instrument & Start time & Mode & Duration \\
           & (UTC)      &      & (s)      \\
\hline
pn   & 2012-12-20T19:45:17 & SW   & 87\,277 \\ 
MOS1 & 2012-12-20T19:40:02 & FF   & 87\,532 \\ 
MOS2 & 2012-12-20T19:40:02 & FF   & 87\,520 \\
\hline
\end{tabular}
\tablefoot{The EPIC cameras were operated in imaging mode and the thin filter was used (\textsf{obsid} 0691970101).}
\end{table}

We re-observed \jten\ with \xmm\ two years later to further improve the characterisation of the source's spectral energy distribution, confirm the candidate spin period, and eventually constrain the pulsar spin-down. We report here the results of this observation in detail. We also reanalysed the AO9 dataset to consistently compare our findings in a joined timing and spectral X-ray analysis. The paper's outline is as follows: in Section~\ref{sec_xmmdatared} we describe the new \xmm\ observation and the data reduction. Analysis and results are presented in detail in Section~\ref{sec_xmmanalysis}. We examine the evolutionary state and possible nature of the neutron star in light of the observed properties of the known galactic INS population in Section~\ref{sec_discussion}. The main conclusions and results are summarised in Section~\ref{sec_summary}.
\section{\xmm\ observation and data reduction\label{sec_xmmdatared}}
For details on the instrumentation set-up and execution of the AO9 observation, we refer to \citet{pir12}. The new \xmm\ observation of \jten\ (AO11) was carried out on 2012 December 20 for a total exposure time of 87.714\,ks. 
Table~\ref{tab_exposureinfoEPIC} contains information on the scientific exposures and instrumentation set-up of the EPIC pn \citep{str01} and MOS \citep{tur01} detectors.
We adopted the thin filter for all EPIC instruments, given its better response at soft X-ray energies and the very low ($<1\%$) expected level of spectral pile-up at aimpoint.
The pn camera was operated in small window (SW) mode, providing high time resolution ($6$\,ms) at the expense of a reduced exposure (deadtime of $\sim29\%$). Although the MOS cameras operating in timing uncompressed mode could provide sufficient resolution (1.75\,ms) for investigating the periodicity at $P\sim18.6$\,ms, calibration uncertainties between the EPIC detectors \citep[e.g.][]{zan11} and the impossibility of using the data for spectral analysis or for astrometrical corrections motivated us to choose the full-frame (FF) imaging mode instead. 
Standard data reduction was performed with SAS~13 (\textsf{\small xmmsas\_20131209\_1901-13.5.0}), using the latest calibration files. We processed the MOS and pn raw event files using the EPIC meta tasks \textsf{\small emchain} and \textsf{\small epchain}, respectively, applying default corrections. We ensured that the pn event file was clean of unrecognised time jumps (i.e.~those uncorrected by standard SAS processing) by checking the output files of the \textsf{\small epproc} routine. 
Additionally, we reprocessed and reduced the AO9 observation consistent with the new dataset.
\begin{table*}[t]
\caption{Parameters of \jtenfull, as extracted from the AO11 (and AO9) \xmm\ EPIC images
\label{tab_sourceMLparam}}
\centering
\begin{tabular}{l r r r r r}
\hline\hline
Parameter               & pn                                & MOS1                              & MOS2                              & EPIC                    & EPIC (AO9) \\
\hline
Detection likelihood    & $6596$                            & $1613$                            & $2280$                            & $10567$                 & $8952$ \\
Counts                  & $5.54(9)\times10^3$               & $1.55(5)\times10^3$               & $1.71(5)\times10^3$               & $8.85(12)\times10^3$    & $7.83(11)\times10^3$ \\
Counts ($0.2-0.5$\,keV) & $1.11(4)\times10^3$               & $2.70(22)\times10^2$              & $2.25(18)\times10^2$              & $1.61(5)\times10^3$     & $1.53(5)\times10^3$ \\
Counts ($0.5-1.0$\,keV & $3.50(7)\times10^3$               & $8.3(4)\times10^2$                & $1.00(3)\times10^3$               & $5.36(9)\times10^3$     & $4.73(9)\times10^3$ \\
Counts ($1.0-2.0$\,keV) & $9.2(4)\times10^2$                & $4.44(29)\times10^2$              & $4.81(27)\times10^2$              & $1.86(5)\times10^3$     & $1.56(5)\times10^3$ \\
Counts ($2.0-4.5$\,keV) & $11\pm11$                         & $0\pm2.0$                         & $0.0\pm2.0$                       & $13\pm11$               & $0\pm4$ \\
Counts ($4.5-12$\,keV)  & $2\pm6$                           & $0\pm2.0$                         & $4\pm4$                           & $5\pm7$                 & $12\pm10$ \\
Rate ($10^{-2}$\,s$^{-1}$) & $9.18(15)$                     & $2.23(8)$                         & $2.18(6)$                         & $13.67(18)$             & $13.27(19)$ \\
HR$_1$                  & $0.518\pm0.015$                   & $0.51\pm0.03$                     & $0.633\pm0.027$                   & $0.539\pm0.012$         & $0.511\pm0.013$ \\
HR$_2$                  & $-0.585\pm0.015$                  & $-0.31\pm0.04$                    & $-0.35\pm0.03$                    & $-0.485\pm0.012$        & $-0.505\pm0.014$ \\
HR$_3$                  & $-0.976\pm0.023$                  & $-1.000\pm0.007$                  & $-1.000\pm0.010$                  & $-0.986\pm0.012$        & $-1.000\pm0.006$ \\   
\hline
RA                      & $10$\ \ $46$\ \ $08.75(15)$       & $10$\ \ $46$\ \ $08.77(26)$       & $10$\ \ $46$\ \ $08.76(21)$       & $10$\ \ $46$\ \ $08.7(6)$\,$^{\dagger}$ & $10$\ \ $46$\ \ $08.8(4)$\,$^{\dagger}$ \\
DEC                     & $-59$\ \ $43$\ \ $05.32(15)$      & $-59$\ \ $43$\ \ $04.97(26)$      & $-59$\ \ $43$\ \ $05.07(21)$      & $-59$\ \ $43$\ \ $06.7(5)$\,$^{\dagger}$ & $-59$\ \ $43$\ \ $06.8(5)$\,$^{\dagger}$ \\
\hline
\end{tabular}
\tablefoot{Counts and rates are given in the total \xmm\ energy band ($0.2-12$\,keV), unless otherwise specified. $^\dagger$The marked EPIC source coordinates are astrometrically corrected, using as reference the 2MASS catalogue (see text). The corresponding $1\sigma$ errors take the astrometric errors into account.}
\end{table*}

A few minor background flares were registered during the observation, effectively reducing the exposures in the MOS1 and MOS2 cameras to 79\,ks and 84\,ks, respectively. In the pn camera, the background level for the duration of the observation was below what is recommended by standard filtering of `good-time intervals'. Therefore, the effective observing time (60\,ks) was only reduced by the camera livetime in SW mode.

For the analysis, we filtered the event lists to exclude bad CCD pixels and columns, as well as to retain the predefined photon patterns with the highest quality energy calibration. Unless otherwise noted, single and double events were selected for pn (pattern $\le4$) and single, double, triple, and quadruple for MOS (pattern $\le12$). We defined the source centroid for each EPIC camera with the task \textsf{\small eregionanalyse}, which results at optimal radii of $\sim20''$ ($0.3-2$\,keV).
Background circular regions of size $60''$ to $100''$ (depending on the camera) were defined off-source on the same CCD as the target whenever possible. 

The detected source count rates, hardness ratios, and other parameters based on a maximum likelihood PSF fitting are listed in Table~\ref{tab_sourceMLparam}. These are determined with the SAS task \textsf{\small emldetect} on images created for each camera and in each of the five predefined \xmm\ energy bands. For comparison, we also list the source parameters as determined from the EPIC detection in the AO9 observation. The errors are nominal $1\sigma$ statistical uncertainties. There is no sign of variability in the source's overall properties; typically, \jten\ displays a very soft energy distribution, with no counts above $2$\,keV, and count rates and hardness ratios are consistent within errors between the two \xmm\ observations (see also Sect.~\ref{sec_spectralanalysis}).
\begin{figure}
\begin{center}
\includegraphics*[width=0.495\textwidth]{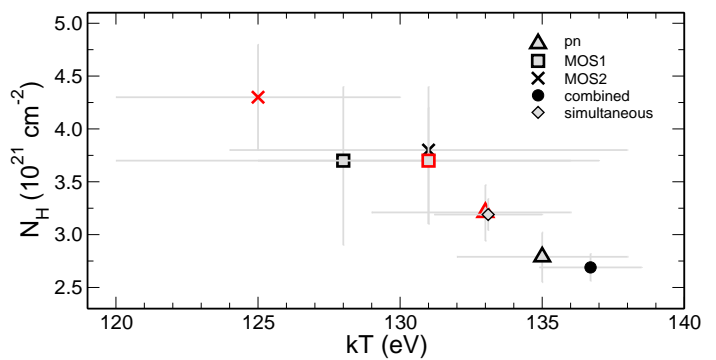}
\end{center}
\caption{Best-fit column density $\nh$ as a function of blackbody temperature $kT$, for each individual (pn/MOS1/MOS2) spectrum used in the analysis (see legend). Also shown are the results by fitting all data simultaneously, as well as results of the stacked (combined) spectrum. AO11 datasets are highlighted in red. Error bars are $1\sigma$ confidence levels.\label{fig_NhkT_obscam}}
\end{figure}

We used the task \textsf{\small eposcorr} to refine the astrometry, by cross-correlating the list of EPIC X-ray source positions with those of near-infrared (2MASS, \citealt{skr06}) objects lying within $15'$ from \jten. 
Based on a number of 60 X-ray/infrared matches, we found a slight offset in right ascension, not significant within the astrometric errors, and an offset of $1.6''\pm0.5''$ in declination. We double-checked the offsets with \textsf{\small eposcorr} using other catalogues, in particular the \chan\ Carina Cluster Project catalogue of X-ray sources (CCCP; \citealt{bro11}), the Guide Star Catalog (GSC~2.3.2; \citealt{las08}), and the Naval Observatory Merged Astrometric Dataset (NOMAD; \citealt{zac04}). The EPIC source position (Table~\ref{tab_sourceMLparam}), updated accordingly, was found to be consistent within errors with previous determinations. For consistency, we revised the source position in the AO9 observation also by cross-correlating the EPIC source list with the 2MASS catalogue. We found offsets of $-2.6''\pm0.4''$ in right ascension and $1.6''\pm0.5''$ in declination (46 matches). The updated source coordinates in the AO9 epoch are also listed in Table~\ref{tab_sourceMLparam}.

\begin{table*}[t]
\caption{Best-fit spectral parameters and count rates of \jtenfull\ per \xmm\ observation and EPIC camera\label{tab_NhkT_obscam}}
\centering
\begin{tabular}{l l l l l l l l l l l l l}
\hline\hline
Ref. & MJD & \multicolumn{3}{c}{$\nh$ ($\times10^{21}$\,cm$^{-2}$)} & & \multicolumn{3}{c}{$kT$ (eV)} & & \multicolumn{3}{c}{Net count rate ($\times10^{-2}$\,s$^{-1}$)}\\
\cline{3-5}\cline{7-9}\cline{11-13}
     & (days) & pn & MOS1 & MOS2 & & pn & MOS1 & MOS2 & & pn & MOS1 & MOS2 \\ 
\hline
AO9  & 55536.531753 & $2.79_{-0.23}^{+0.24}$ & $3.7_{-0.7}^{+0.8}$ & $3.8_{-0.6}^{+0.7}$ & & $135(3)$ & $128(8)$ & $131(7)$ & & $7.39(12)$ & $1.54(6)$ & $1.66(6)$ \\
AO11 & 56282.326597 & $3.21_{-0.26}^{+0.27}$ & $3.7_{-0.5}^{+0.6}$ & $4.3(5)$ & & $133_{-3}^{+4}$ & $131(6)$ & $125(5)$ & & $7.39(12)$ & $1.63(5)$ & $1.73(5)$ \\
\hline
\multicolumn{2}{l}{Simultaneous (per camera)} & $2.98(18)$ & $3.7_{-0.4}^{+0.5}$ & $4.1(4)$ & & $134.2(2.4)$ & $130(5)$ & $127(4)$ & & $7.39(17)$ & $1.58(8)$ & $1.70(8)$ \\
\multicolumn{2}{l}{Simultaneous (EPIC)} & \multicolumn{3}{c}{$3.20(15)$} & & \multicolumn{3}{c}{$133.1(1.9)$} & & \multicolumn{3}{c}{$21.34(20)$} \\
\multicolumn{2}{l}{Combined} & \multicolumn{3}{c}{$2.69(13)$} & & \multicolumn{3}{c}{$136.7(1.8)$} & & \multicolumn{3}{c}{$21.94(19)$} \\
\hline
\end{tabular}
\tablefoot{Errors are $1\sigma$ confidence levels. The model fitted to the data is a simple absorbed blackbody (\textsf{tbabs$\ast$bbody}). Reduced $\chi^2_\nu$ values are within 0.9 and 3.4, for 23 to 30 degrees of freedom (individual fits). Count rates in energy band $0.3-12$\,keV are subtracted for the background.} 
\end{table*}
The statistics for the EPIC lightcurves (corrected for bad pixels, deadtime, exposure, as well as background counts, and binned into 875\,s intervals) show the $3\sigma$ upper limits for the r.m.s. fractional variation of 0.12, 0.11 and 0.24 for pn, MOS1, and MOS2, respectively. 
On the basis of the same lightcurves, the reduced chi-square, assuming a constant flux, is 1.02 (pn, 99 d.o.f.), 0.97, and 0.71 (MOS1 and MOS2; 98 d.o.f.). 
\section{Analysis and results\label{sec_xmmanalysis}}
\subsection{Spectral analysis\label{sec_spectralanalysis}}
The analysis of the EPIC data is based on source and background spectra extracted from regions as described in Section~\ref{sec_xmmdatared}, together with the respective response matrices and ancillary files created for each of the EPIC cameras. Owing to the frequent readout of pn in SW mode, detector noise dominates the low-energy count distribution. We therefore restricted the analysis of the pn camera to photons with energies above 0.35\,keV. The energy band of the MOS cameras is restricted above 0.3\,keV, which is in accordance with the guidelines and calibration status of the instrument. 
Including the AO9 observation, the available data amount to a total of six spectra and $\sim1.7\times10^4$ counts ($0.3-12$\,keV), of which $10\%$ can be ascribed to background. 
For each spectrum, energy channels within 0.3/0.35\,keV and $2$\,keV were rebinned according to a minimum number of 25 counts per spectral bin. We also took care not to oversample the instrument energy resolution at a given bin by more than a factor of three, which is especially important at low energies; this is done by fixing the parameter \textit{oversample = 3} in the SAS task \textsf{\small specgroup}.

To fit the spectra we used XSPEC~12.8.2 \citep{arn96}. Unless otherwise noted, the fit parameters were allowed to vary freely and within reasonable ranges. The photoelectric absorption model and elemental abundances of \citet[][\textsf{\small tbabs} in XSPEC]{wil00} were adopted to account for the interstellar absorption. Owing to the intrinsically soft energy distribution of \jten\ and its location in the galactic plane ($b=-0.6^\circ$), as well as to uncertainties resulting from the choice of abundance table in XSPEC, we estimate that the amount of intervening material derived from the spectral fits is uncertain by $15\%$. 

\begin{figure}
\begin{center}
\includegraphics*[width=0.495\textwidth]{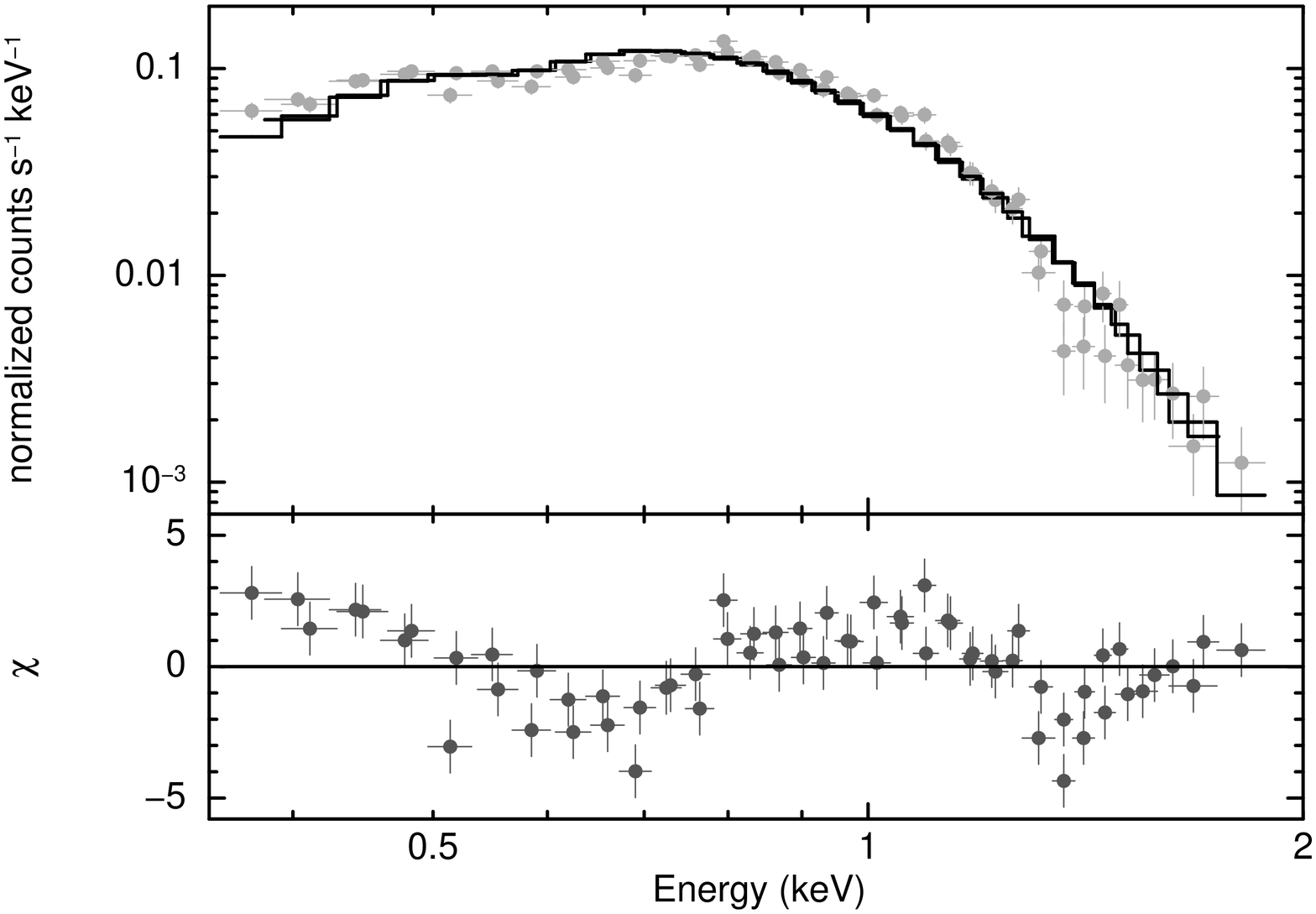}\\[0.25cm]
\includegraphics*[width=0.495\textwidth]{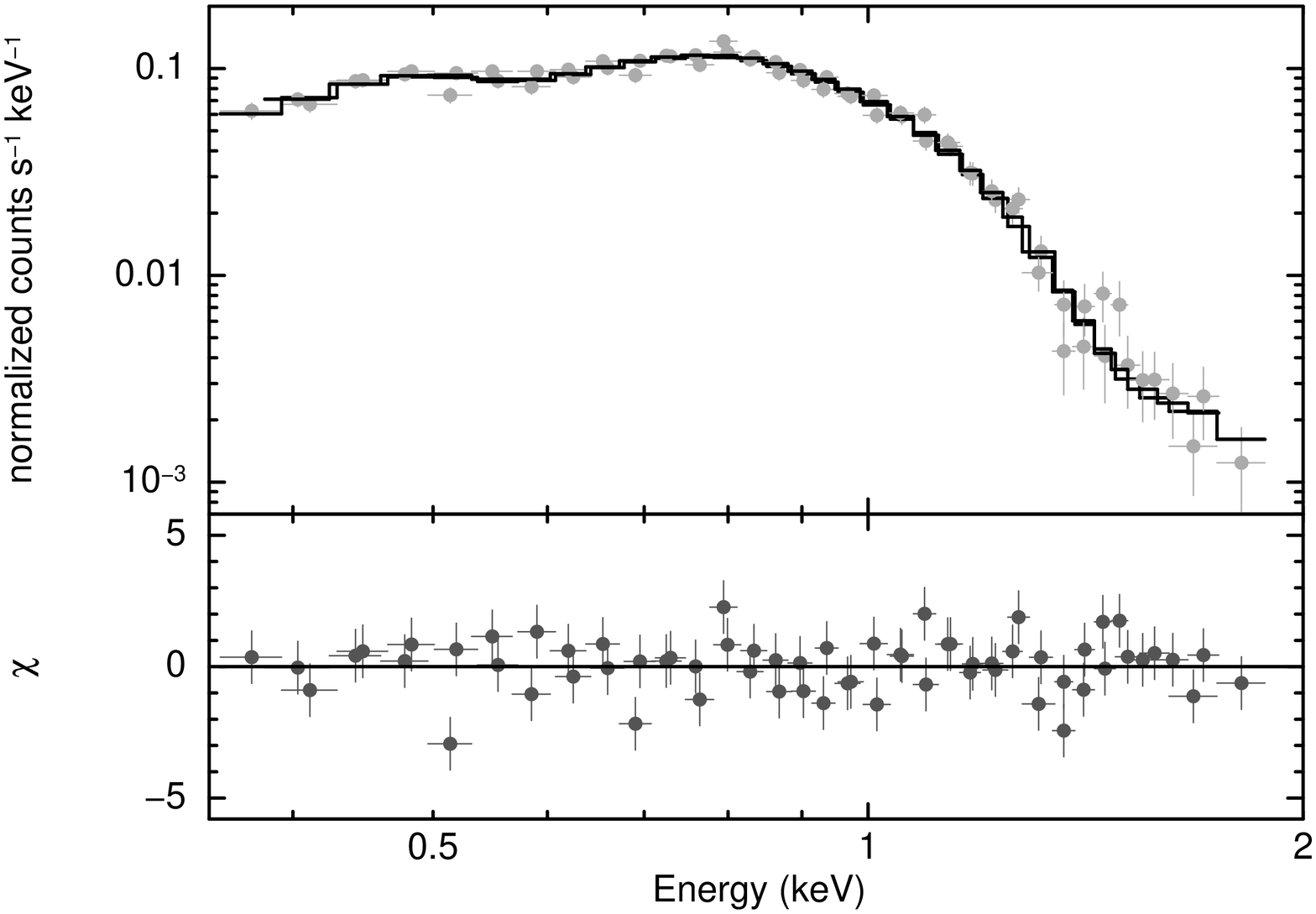}
\end{center}
\caption{Results of spectral fitting of source \jtenfull. We show the two pn spectra and folded best-fit models, with residuals. \textit{Top:} Simple absorbed blackbody. \textit{Bottom:} Best-fit neutron star atmosphere at $d\sim2$\,kpc ($B=10^{12}$\,G, $T_{\rm eff}\sim10^6$\,K, $\nh=2.6\times10^{21}$\,cm$^{-2}$). Two Gaussian lines in absorption are added to the model (with energies around 0.55\,keV and 1.32\,keV).\label{fig_bestBBfit}}
\end{figure}
\begin{table*}[t]
\caption{Results of spectral fitting of blackbody models
\label{tab_resultspecBB}}
\centering
\begin{tabular}{l r r c c c c r c r}
\hline\hline
Model & $\chi^2_{\nu}$\,(d.o.f.) & \multicolumn{1}{c}{$\nh$} & \multicolumn{1}{c}{$kT^\infty_1$} & \multicolumn{1}{c}{$kT^\infty_2$}  & \multicolumn{1}{c}{$\epsilon_1$} & \multicolumn{1}{c}{$\epsilon_2$} & \multicolumn{1}{c}{$F_{\rm X}$\,\tablefootmark{$\ddag$}} & \multicolumn{1}{c}{$R^\infty_1$\,\tablefootmark{$\dagger$}} & \multicolumn{1}{c}{$R^\infty_2$\,\tablefootmark{$\dagger$}} \\ 
 & & ($\times10^{21}$\,cm$^{-2}$) & \multicolumn{1}{c}{(eV)} & \multicolumn{1}{c}{(eV)} & \multicolumn{1}{c}{(keV)} & \multicolumn{1}{c}{(keV)} & (\,erg\,s$^{-1}$\,cm$^{-2}$) & \multicolumn{1}{c}{(km)} & \multicolumn{1}{c}{(km)}\\
\hline
(A) & 1.98\,(169) & $3.20(15)$ & -- & $133.1(1.9)$ & -- & -- & $9.3^{+1.8}_{-1.5}\times10^{-13}$ & -- & $3.8_{-0.4}^{+0.5}$ \\
(B) & 1.34\,(167) & $5.3(3)$ & $28.6^{+2.3}_{-2.1}$ & $117.8(2.2)$ & -- & -- & $6_{-3}^{+2.4}\times10^{-10}$ & $4_{-2}^{+4}\times10^3$\,\tablefootmark{a} & $6.8^{+0.8}_{-0.7}$ \\
(C) & 1.48\,(168) & $3.75(17)$ & -- & $121.3(2.1)$ & -- & -- & $1.5_{-0.5}^{+0.9}\times10^{-12}$ & -- & $5.8_{-2.2}^{+4}$ \\ 
(D) & 1.08\,(165) & $2.17_{-0.20}^{+0.22}$ & -- & $157(5)$ & -- & $1.289_{-0.017}^{+0.018}$ & $4.6_{-1.4}^{+6}\times10^{-13}$ & -- & $1.9_{-0.4}^{+5}$ \\ 
(E) & 1.03\,(165) & $2.16(20)$ & -- & $144(3)$ & $0.577_{-0.012}^{+0.013}$ & $1.361_{-0.016}^{+0.015}$ & $5.4_{-1.1}^{1.4}\times10^{-13}$ & -- & $2.5(4)$ \\
(F) & 1.02\,(165) & $3.4_{-0.3}^{+0.4}$ & $36(4)$ & $150(5)$ & -- & $1.270_{-0.019}^{+0.018}$ & $1.2_{-0.6}^{+1.9}\times10^{-11}$ & $310^{+320}_{-150}$\,\tablefootmark{b} & $2.4_{-0.3}^{+0.4}$ \\
\hline
\end{tabular}
\tablefoot{
Errors are $1\sigma$ confidence levels. All \textsf{tbabs} models assume \citet{wil00} abundances. Models: (A)~\textsf{tbabs(bbody)}; (B)~\textsf{tbabs(bbody+bbody)}; (C)~\textsf{vphabs(bbody)}, with best-fit oxygen overabundance of $Z_{\rm O}=1.60(7)$ in solar units; (D)~\textsf{vphabs(bbody+gauss)}, with $Z_{\rm O}=1.64(12)$, best-fit Gaussian $\sigma\sim0.2$\,keV, and line equivalent width of $EW\sim220$\,eV; (E)~\textsf{tbabs(bbody+gauss+gauss)}, with fixed Gaussian $\sigma_1=0.1$\,keV and $\sigma_2=0.2$\,keV, and $EW_1\sim90$\,eV and $EW_2\sim120$\,eV; (F)~\textsf{tbabs(bbody+bbody+gauss)}, with fixed Gaussian $\sigma_2=0.2$\,keV and $EW_2\sim210$\,eV. 
\tablefoottext{$\ddag$}{The unabsorbed flux is in energy band $0.2-12$\,keV.}
\tablefoottext{$\dagger$}{The radiation radii are estimated normalising the distance to the source to that of \etacar, $d_{\rm Car}=2.3$\,kpc.}
\tablefoottext{a}{In order to constrain $R_1^\infty=20$\,km, the source must be at 10\,pc ($R_2^\infty\sim2$\,km).
\tablefoottext{b}}{In order to constrain $R_1^\infty=20$\,km, the source must be at 150\,pc ($R_2^\infty\sim2$\,km).}
}
\end{table*}
First, we fitted each of the six spectra individually, using a simple absorbed blackbody model. The resulting best-fit parameters (Table~\ref{tab_NhkT_obscam} and Fig.~\ref{fig_NhkT_obscam}) argue against any significant variability between the two \xmm\ observations. However, the $\nh$ and $kT$ derived from the MOS spectra, even though statistically consistent, are systematically higher and softer, by $30\%$ and $5\%$ respectively, than those derived from pn.
We then proceeded with the analysis using two different approaches, both relying on the assumption that the spectral parameters of \jten\ are constant between the \xmm\ observations: (\textit{i}) we performed simultaneous fits of all spectra, where we allowed for a renormalisation factor to account for cross-calibration uncertainties between the detectors, and (\textit{ii}) using the task \textsf{\small epicspeccombine}, we merged all six spectra and corresponding background and response files into one single, stacked dataset.
While both approaches give a similarly well-constrained parameter space, the exercise shows that the best-fit $\nh$ and $kT$, derived from the second approach, are biased and inconsistent with those from individual fits. On the other hand, the results from the first approach, while naturally dominated by the better statistics of the two pn spectra, are consistent with most measurements.
We henceforth adopted for the spectral analysis simultaneous fits, including all EPIC data from the two \xmm\ observations accordingly.

We summarise the results of the analysis in Tables~\ref{tab_resultspecBB}, \ref{tab_resultspecNSA}, and \ref{tab_resultspecNSMAX}. Table~\ref{tab_resultspecBB} contains the results of spectral fits where we assumed a (single- or double-temperature) blackbody continuum, and tested for different element abundances and for the presence of absorption lines. The fitted models are labelled (A) to (F), as reference for the text (see caption). To test for overabundance of elements with transitions in the range of 0.3\,keV to 2\,keV (C, N, O, Ne, Mg, etc), we adopted the variable photoelectric absorption model $\textsf{\small vphabs}$ and restricted abundance values in the range $1<Z<5$ in solar units.

The best-fit parameters of (more physically-motivated) neutron star atmosphere models are in Tables~\ref{tab_resultspecNSA} and \ref{tab_resultspecNSMAX}, where they are labelled (a) to (e). We tested magnetised and non-magnetised model atmospheres, consisting of fully- and partially-ionized hydrogen, as well as of carbon and other mid-$Z$ elements (the models \textsf{\small nsa}, \textsf{\small nsmaxg}, and \textsf{\small carbatm} in XSPEC; see \citealt{zav96,pav95,hop08,mor07,sul14}). Wherever appropriate, the magnetic field was held fixed at values $B=0,10^{12},10^{13}$\,G (\textsf{\small nsa}, Table~\ref{tab_resultspecNSA}); for \textsf{\small nsmaxg}, we tested 19 models with $B=(0.01-30)\times10^{12}$\,G (only the results with $\chi^2_\nu<2$ are listed in Table~ \ref{tab_resultspecNSMAX}). We first assumed canonical values for the neutron star mass and radius, $M_{\rm ns}=1.4$\,M$_{\odot}$ and $R_{\rm ns}=10$\,km; these parameters were also allowed to vary to check for an improved fit. The size of the emission region, with respect to the neutron star physical radius, can also be parametrised for \textsf{\small nsmaxg} models; we restricted values to have $R_{\rm em}\le R_{\rm ns}$. Finally, we also added a second thermal component, varied the elemental abundances, and added lines to the continuum to obtain statistically acceptable fits and derive plausible physical solutions for the neutron star. For double-temperature \textsf{\small nsa} and \textsf{\small nsmaxg} models, we fixed the mass of the neutron star at 1.4\,M$_\odot$, let the radii ($R_{\rm ns}$ or $R_{\rm em}$) of the two components vary independently of each other, and limited the distance to the source to within 10\,pc and 50\,kpc. The mass, magnetic field, and distance normalisation of the second thermal component were set equal to those of the first.
\begin{sidewaystable*}
\caption{Results of spectral fitting of neutron star atmosphere models (\textsf{nsa})
\label{tab_resultspecNSA}}
\centering
\begin{tabular}{l c c r r r r r r r c c c r r}
\hline\hline
 & $B$ & $\chi^2_\nu$\,(d.o.f.) & \multicolumn{1}{c}{$\nh$} & \multicolumn{1}{c}{$T_{\rm eff, 1}$} & \multicolumn{1}{c}{$T_{\rm eff, 2}$} & \multicolumn{1}{c}{$\epsilon_1$} & $EW_1$ &\multicolumn{1}{c}{$\epsilon_2$} & $EW_2$ & $M_{\rm ns}$ & \multicolumn{1}{c}{$R_{\rm em, 1}^\infty$} & \multicolumn{1}{c}{$R_{\rm em, 2}^{\infty}$} & $\log(L_{\rm X})$ & \multicolumn{1}{c}{$d$} \\
\cline{7-8}\cline{9-10}
 & (G) & & \multicolumn{1}{c}{($10^{21}$\,cm$^{-2})$} & ($10^5$\,K) & ($10^5$\,K) & \multicolumn{2}{c}{(keV)} & \multicolumn{2}{c}{(keV)} & (M$_\odot$) & \multicolumn{1}{c}{(km)} & \multicolumn{1}{c}{(km)} & (erg\,s$^{-1}$) & \multicolumn{1}{c}{(kpc)} \\
\hline
(a) & 0 & 1.92\,(169) & $4.85(11)$ & $4.38_{-0.18}^{+0.21}$ & -- & \multicolumn{2}{c}{--} & \multicolumn{2}{c}{--} & 1.4 & 13 & -- & $31.42^{+0.08}_{-0.07}$ & 0.17 \\
 & 1e12 & 1.63\,(169) & $2.59_{-0.15}^{+0.13}$ & $30.0_{-2.3}^{+4}$ & -- & \multicolumn{2}{c}{--} & \multicolumn{2}{c}{--}  & 2 & 18 & -- & $34.44^{+0.22}_{-0.14}$ & $8.0^{+1.8}_{-6}$ \\
 & 1e13 & 1.81\,(169) & $3.70_{-0.23}^{+0.16}$ & $21.2_{-2.9}^{+1.3}$ & -- & \multicolumn{2}{c}{--} & \multicolumn{2}{c}{--}  & 2.5 & 25 & -- & $34.00^{+0.11}_{-0.26}$ & $2.58(4)$ \\
(b) & 0 & 2.19\,(166) & $4.79_{-0.18}^{+0.10}$ & $3.00_{-0.21}^{+0.12}$ & $10$ & \multicolumn{2}{c}{--} & \multicolumn{2}{c}{--}  & 1.4\,\tablefootmark{$\star$} & $22.5_{-0.9}^{+0.6}$ & $6$ & $32.34_{-1.1}^{+0.05}$ & $0.335_{-0.03}^{+0.004}$ \\ 
 & 1e12 & 1.68\,(166) & $2.84_{-0.14}^{+0.20}$ & $4.4_{-1.7}^{+2.7}$ & $24.58_{-0.5}^{+0.24}$ & \multicolumn{2}{c}{--} & \multicolumn{2}{c}{--}  & 1.4\,\tablefootmark{$\star$} & 23 & 6 & $33.82_{-0.04}^{+0.06}$ & $3.85_{-0.04}^{+0.08}$ \\ 
 & 1e13 & 1.96\,(166) & $4.71_{-0.26}^{+0.17}$ & $3.1_{-0.5}^{+0.8}$ & $13.52_{-0.12}^{+0.13}$ & \multicolumn{2}{c}{--} & \multicolumn{2}{c}{--}  & 1.4\,\tablefootmark{$\star$} & 23 & 6 & $32.794_{-0.024}^{+0.04}$ & $0.509_{-0.005}^{+0.010}$ \\
(c) & 0 & 1.57\,(165) & $4.98(11)$ & $4.41_{-0.22}^{+0.26}$ & -- & \multicolumn{2}{c}{--} & \multicolumn{2}{c}{--}  & 1.5 & 13 & -- & $31.39_{-0.09}^{+0.10}$ & 0.12 \\
 & 1e12 & 1.32\,(165) & $2.78_{-0.20}^{+0.13}$ & $33_{-4}^{+3}$ & -- & \multicolumn{2}{c}{--} & \multicolumn{2}{c}{--}  & 1.9 & 18 & -- & $34.53_{-0.26}^{+0.15}$ & $7_{-5}^{+3}$ \\
 & 1e13 & 1.89\,(165) & $6.34_{-0.3}^{+0.19}$ & $4.8_{-0.4}^{+1.3}$ & -- & \multicolumn{2}{c}{--} & \multicolumn{2}{c}{--}  & 0.8 & 11 & -- & $31.5_{-0.6}^{+0.4}$ & 0.10 \\
(d) & 0 & 1.39\,(166) & $2.80(3)$ & $4.3_{-1.8}^{+1.5}$ & $15.1_{-4}^{+2.7}$ & \multicolumn{2}{c}{--} & $1.346(15)$ & $0.3$ & 1.4\,\tablefootmark{$\star$} & 23 & $6.099_{-0.016}^{+0.025}$ & $33.1_{-0.5}^{+0.3}$ & $1.93_{-0.23}^{+0.25}$ \\
 & 1e12 & 1.15\,(166) & $1.49_{-0.13}^{+0.22}$ & $6(4)$ & $38_{-4}^{+9}$ & \multicolumn{2}{c}{--} & $1.209_{-0.014}^{+0.012}$ & $0.25$ & 1.4\,\tablefootmark{$\star$} & 23 & 6 & $35(5)$ & $13.8_{-0.4}^{+0.6}$ \\
 & 1e13 & 1.28\,(166) & $2.58_{-0.10}^{+0.23}$ & $4.1_{-0.6}^{+4}$ & $22.6_{-0.4}^{+0.3}$ & \multicolumn{2}{c}{--} & $1.309_{-0.013}^{+0.012}$ & $0.2$ & 1.4\,\tablefootmark{$\star$} & 23 & 6 & $33.68_{-3}^{+0.07}$ & $3.30_{-0.04}^{+0.09}$ \\
(e) & 0 & 1.08\,(165) & $2.9(3)$ & $6.3_{-0.3}^{+0.5}$ & -- & $0.548_{-0.013}^{+0.017}$ & $0.07$ & $1.327_{-0.028}^{+0.019}$ & $0.26$ & 1.4\,\tablefootmark{$\star$} & 13 & -- & $32.05_{-0.10}^{+0.14}$ & $0.654(6)$ \\
 & 1e12 & 1.05\,(165) & $2.59_{-0.3}^{+0.21}$ & $10.1_{-0.4}^{+0.8}$ & -- & $0.549_{-0.014}^{+0.013}$ & $0.07$ & $1.320_{-0.014}^{+0.017}$ & $0.22$ & 1.4\,\tablefootmark{$\star$} & 13 & -- & $32.87_{-0.07}^{+0.13}$ & $1.997(19)$ \\
 & 1e13 & 1.12\,(165) & $2.91_{-0.4}^{+0.20}$ & $9.4_{-0.5}^{+0.4}$ & -- & $0.545_{-0.011}^{+0.013}$ & $0.08$ & $1.352_{-0.016}^{+0.014}$ & $0.28$ & 1.4\,\tablefootmark{$\star$} & 13 & -- & $32.75_{-0.09}^{+0.08}$ & $1.524_{-0.014}^{+0.015}$ \\
\hline
\end{tabular}
\tablefoot{Errors are $1\sigma$ confidence levels (unconstrained parameters are shown with no errors for illustrative purposes). All \textsf{tbabs} models assume \citet{wil00} abundances. All listed models consist of hydrogen atmospheres. Models: (a)~\textsf{tbabs(nsa)}; (b)~\textsf{tbabs(nsa+nsa)}; (c)~\textsf{vphabs(nsa)}, with best-fit oxygen overabundance of $Z_{\rm O}\sim1.3-1.7$ in solar units; (d)~\textsf{tbabs(nsa+nsa+gauss)}, with fixed Gaussian $\sigma_2=0.2$\,keV; (e)~\textsf{tbabs(nsa+gauss+gauss)}, with fixed Gaussian $\sigma_1=0.1$\,keV and $\sigma_2=0.2$\,keV. 
\tablefoottext{$\star$}{Parameter held fixed during fitting.}
}
\end{sidewaystable*}
\begin{sidewaystable*}
\caption{Results of spectral fitting of neutron star atmosphere models (\textsf{nsmaxg})
\label{tab_resultspecNSMAX}}
\centering
\begin{tabular}{l c c r r r r r r r c c c r r}
\hline\hline
 & $B$ & $\chi^2_\nu$\,(d.o.f.) & \multicolumn{1}{c}{$\nh$} & \multicolumn{1}{c}{$T_{\rm eff, 1}$} & \multicolumn{1}{c}{$T_{\rm eff, 2}$} & \multicolumn{1}{c}{$\epsilon_1$} & $EW_1$ &\multicolumn{1}{c}{$\epsilon_2$} & $EW_2$ & $M_{\rm ns}$ & \multicolumn{1}{c}{$R_{\rm em, 1}^\infty$} & \multicolumn{1}{c}{$R_{\rm em, 2}^{\infty}$} & $\log(L_{\rm X})$ & \multicolumn{1}{c}{$d$} \\
\cline{7-8}\cline{9-10}
 & (G) & & \multicolumn{1}{c}{($10^{21}$\,cm$^{-2})$} & ($10^5$\,K) & ($10^5$\,K) & \multicolumn{2}{c}{(keV)} & \multicolumn{2}{c}{(keV)} & (M$_\odot$) & \multicolumn{1}{c}{(km)} & \multicolumn{1}{c}{(km)} & (erg\,s$^{-1}$) & \multicolumn{1}{c}{(kpc)} \\
\hline
(a) & 2e13 & 1.64\,(166) & $7.14^{+0.05}_{-0.06}$ & $3.98^{+0.06}_{-4}$ & -- & \multicolumn{2}{c}{--} & \multicolumn{2}{c}{--}  & 1.1 & 23 & -- & $31.92_{-22}^{+0.03}$ & $0.06$ \\
 & 3e13 & 1.83\,(168) & $5.32^{+0.06}_{-0.05}$ & $5.00^{+0.05}_{-5}$ & -- & \multicolumn{2}{c}{--} & \multicolumn{2}{c}{--}  & 1.4 & 13 & -- & $31.65_{-23}^{+0.02}$ & $0.06$ \\
(b) & 3e13 & 1.86\,(166) & $5.30_{-0.09}^{+0.07}$ & $3.2$ & $79.4$ & \multicolumn{2}{c}{--} & \multicolumn{2}{c}{--}  & 1.4\,\tablefootmark{$\star$} & 13 & $\ll1$ & $30.85$ & $0.051_{-0.023}^{+0.014}$ \\
(c) & 1e10 & 1.69\,(167) & $5.46(21)$ & $3.88_{-0.15}^{+0.16}$ & -- & \multicolumn{2}{c}{--} & \multicolumn{2}{c}{--}  & 1.4\,\tablefootmark{$\star$} & 13 & -- & $31.21(7)$ & 0.09 \\ 
 & 7e12 & 1.92\,(167) & $7.03_{-0.3}^{+0.22}$ & $5.13_{-0.14}^{+0.21}$ & -- & \multicolumn{2}{c}{--} & \multicolumn{2}{c}{--}  & 1.4\,\tablefootmark{$\star$} & 13 & -- & $31.69_{-0.05}^{+0.07}$ & 0.06 \\ 
 & 1e13 & 1.89\,(167) & $6.77_{-0.29}^{+0.20}$ & $5.40_{-0.15}^{+0.20}$ & -- & \multicolumn{2}{c}{--} & \multicolumn{2}{c}{--}  & 1.4\,\tablefootmark{$\star$} & 13 & -- & $31.78_{-0.06}^{+0.05}$ & 0.08 \\ 
 & 3e13 & 1.56\,(167) & $5.15(5)$ & $3_{-3}^{+7}$ & -- & \multicolumn{2}{c}{--} & \multicolumn{2}{c}{--}  & 1.4\,\tablefootmark{$\star$} & 13 & -- & $30.9_{-2.2}^{+2.0}$ & 0.06 \\ 
(d) & 1e10 & 1.34\,(164) & $2.69_{-0.23}^{+0.10}$ & $8.59_{-0.7}^{+0.16}$ & $10$ & \multicolumn{2}{c}{--} & $1.360_{-0.012}^{+0.013}$ & $0.3$ & 1.4\,\tablefootmark{$\star$} & $7.7(2.8)$ & $<1.8$ & $32.1_{-0.5}^{+0.8}$ & $1.001(8)$ \\
 & 1e12 & 1.70\,(164) & $2.80_{-0.05}^{+0.07}$ & $3$ & $10$ & \multicolumn{2}{c}{--} & $1.391_{-0.009}^{+0.008}$ & $0.4$ & 1.4\,\tablefootmark{$\star$} & $\ll2$ & $2.6$ & $31(24)$ & $0.310(3)$ \\
 & 1.3e12 & 1.58\,(164) & $3.72_{-0.27}^{+0.13}$ & $8.36_{-0.11}^{+0.13}$ & $10$ & \multicolumn{2}{c}{--} & $1.369_{-0.011}^{+0.010}$ & $0.3$ & 1.4\,\tablefootmark{$\star$} & $13(4)$ & $<2.0$ & $32.54_{-0.3}^{+0.04}$ & $0.975(8)$ \\
 & 2e12 & 1.58\,(164) & $3.78_{-0.4}^{+0.15}$ & $9$ & $10$ & \multicolumn{2}{c}{--} & $1.382_{-0.011}^{+0.015}$ & $0.3$ & 1.4\,\tablefootmark{$\star$} & $12.4_{-11}^{+0.6}$ & $<2.3$ & $32.57_{-30}^{+0.09}$ & $0.998(3)$ \\
 & 4e12 & 1.47\,(164) & $3.76_{-0.15}^{+0.17}$ & $3$ & $10$ & \multicolumn{2}{c}{--} & $1.382(10)$ & $0.3$ & 1.4\,\tablefootmark{$\star$} & $8.6_{-1.4}^{+0.6}$ & $0.243_{-0.016}^{+0.019}$ & $30.52_{-22}^{+0.13}$ & $0.03$ \\
 & 7e12 & 1.42\,(164) & $3.95(15)$ & $3$ & $10$ & \multicolumn{2}{c}{--} & $1.353_{-0.010}^{+0.011}$ & $0.3$ & 1.4\,\tablefootmark{$\star$} & $12.88_{-11}^{+0.17}$ & $0.364_{-0.022}^{+0.024}$ & $30.87_{-24}^{+0.08}$ & $0.05$ \\
 & 1e13 & 1.41\,(164) & $4.02_{-0.15}^{+0.17}$ & $3$ & $10$ & \multicolumn{2}{c}{--} & $1.350(11)$ & $0.3$ & 1.4\,\tablefootmark{$\star$} & $10.0_{-1.3}^{+0.6}$ & $0.254(16)$ & $30.65_{-22}^{+0.11}$ & $0.03$ \\
 & 2e13 & 1.37\,(164) & $4.15_{-0.25}^{+0.28}$ & $3$ & $10$ & \multicolumn{2}{c}{--} & $1.346_{-0.011}^{+0.012}$ & $0.3$ & 1.4\,\tablefootmark{$\star$} & $11.7(6)$ & $0.446_{-0.05}^{+0.024}$ & $30.82_{-0.10}^{+0.05}$ & $0.06$ \\
 & 3e13 & 1.40\,(164) & $3.90_{-0.25}^{+0.24}$ & $3$ & $10$ & \multicolumn{2}{c}{--} & $1.333_{-0.013}^{+0.015}$ & $0.28$ & 1.4\,\tablefootmark{$\star$} & $12.4_{-0.8}^{+0.7}$ & $0.99_{-0.16}^{+0.10}$ & $31.02_{-0.27}^{+0.11}$ & $0.15$ \\
(e) & 1e10 & 1.05\,(164) & $2.49_{-0.27}^{+0.25}$ & $7.7_{-0.5}^{+0.6}$ & -- & $0.554(14)$ & $0.07$ & $1.349(16)$ & $0.27$ & 1.4\,\tablefootmark{$\star$} & 9 & -- & $32.06_{-0.11}^{+0.12}$ & $0.844(7)$ \\
 & 1e12 & 1.21\,(164) & $3.29_{-0.29}^{+0.4}$ & $7.4(5)$ & -- & $0.525_{-0.009}^{+0.012}$ & $0.08$ & $1.368_{-0.014}^{+0.012}$ & $0.3$ & 1.4\,\tablefootmark{$\star$} & 11 & -- & $32.20_{-0.13}^{+0.11}$ & $0.579(5)$ \\
 & 1.3e12 & 1.20\,(164) & $3.21_{-0.26}^{+0.3}$ & $7.9(5)$ & -- & $0.528_{-0.009}^{+0.006}$ & $0.08$ & $1.366_{-0.013}^{+0.012}$ & $0.3$ & 1.4\,\tablefootmark{$\star$} & 11 & -- & $32.30_{-0.11}^{+0.10}$ & $0.702(6)$ \\
 & 2e12 & 1.19\,(164) & $3.17_{-0.4}^{+0.21}$ & $8.4_{-0.6}^{+0.4}$ & -- & $0.532_{-0.010}^{+0.013}$ & $0.08$ & $1.368_{-0.016}^{+0.013}$ & $0.3$ & 1.4\,\tablefootmark{$\star$} & 10 & -- & $32.37_{-0.12}^{+0.09}$ & $0.793_{-0.006}^{+0.007}$ \\
 & 4e12 & 1.16\,(164) & $3.05_{-0.04}^{+0.21}$ & $9.2_{-0.5}^{+0.4}$ & -- & $0.537_{-0.010}^{+0.013}$ & $0.08$ & $1.357_{-0.015}^{+0.013}$ & $0.29$ & 1.4\,\tablefootmark{$\star$} & 9 & -- & $32.35_{-0.11}^{+0.08}$ & $0.855(7)$ \\
 & 7e12 & 1.15\,(164) & $3.04_{-0.4}^{+0.20}$ & $9.4(5)$ & -- & $0.539_{-0.010}^{+0.013}$ & $0.08$ & $1.353_{-0.015}^{+0.014}$ & $0.28$ & 1.4\,\tablefootmark{$\star$} & 10 & -- & $32.49_{-0.09}^{+0.08}$ & $1.032_{-0.008}^{+0.009}$ \\
 & 1e13 & 1.14\,(164) & $3.03_{-0.4}^{+0.20}$ & $9.5(5)$ & -- & $0.541_{-0.010}^{+0.013}$ & $0.08$ & $1.352_{-0.015}^{+0.014}$ & $0.28$ & 1.4\,\tablefootmark{$\star$} & 9 & -- & $32.47_{-0.09}^{+0.08}$ & $1.017_{-0.008}^{+0.009}$ \\
 & 2e13 & 1.13\,(164) & $2.99_{-0.4}^{+0.21}$ & $9.6(5)$ & -- & $0.545_{-0.012}^{+0.015}$ & $0.07$ & $1.345_{-0.016}^{+0.015}$ & $0.27$ & 1.4\,\tablefootmark{$\star$} & 8.5 & -- & $32.41_{-0.09}^{+0.08}$ & $0.996(8)$ \\
 & 3e13 & 1.08\,(164) & $3.21_{-0.3}^{+0.21}$ & $8.6_{-0.4}^{+0.7}$ & -- & $0.550(11)$ & $0.08$ & $1.313_{-0.017}^{+0.021}$ & $0.20$ & 1.4\,\tablefootmark{$\star$} & 12 & -- & $32.51_{-0.08}^{+0.13}$ & $0.956(8)$ \\
\hline
\end{tabular}
\tablefoot{Errors are $1\sigma$ confidence levels (unconstrained parameters are shown with no errors for illustrative purposes). All \textsf{tbabs} models assume \citet{wil00} abundances. Only results of models with $\chi^2_\nu<2$ are shown. All listed models consist of hydrogen atmospheres. Models: (a)~\textsf{tbabs(nsmaxg)}; (b)~\textsf{tbabs(nsmaxg+nsmaxg)}; (c)~\textsf{vphabs(nsmaxg)}, with $Z_{\rm O}\sim(1.2-1.3)$; (d)~\textsf{tbabs(nsmaxg+nsmaxg+gauss)}, with fixed Gaussian $\sigma_2=0.2$\,keV; (e)~\textsf{tbabs(nsmaxg+gauss+gauss)}, with fixed Gaussian $\sigma_1=0.1$\,keV and $\sigma_2=0.2$\,keV. \tablefoottext{$\star$}{Parameter held fixed during fitting.}}
\end{sidewaystable*}

The analysis confirms the results of \citet{pir12}, with much improved count statistics. Overall, the spectrum of the source is soft and purely thermal. However, significant deviations from a Planckian shape are identified around energies $0.55$\,keV and $1.35$\,keV (see top plot of Fig.~\ref{fig_bestBBfit}). The residuals are present independently of the choice of the thermal continuum, abundance table, source or background extraction regions, as well as regardless of the EPIC instrument and \xmm\ observation.

The best-fit absorbed blackbody (A) has $kT=(133.1\pm1.9)$\,eV and $\nh=(3.20\pm0.15)\times10^{21}$\,cm$^{-2}$, with $\chi^2_\nu=1.981$ for 169 degrees of freedom (d.o.f.) (Table~\ref{tab_resultspecBB}). The column density is consistent with the one towards Eta Carina, $\nh\sim3\times10^{21}$\,cm$^{-2}$ \citep{leu03}. Assuming the neutron star is at a comparable distance, $d_{\rm Car}=2.3$\,kpc \citep{smi06a}, the redshifted radiation radius derived from the blackbody fit is $R_\infty=(3.8_{-0.4}^{+0.5})(d/d_{\rm Car})^{-1}$\,km. We note that the redshifted radiation radii of the \msev, as derived from X-ray blackbody fits and distance estimates, are in the range of roughly 2\,km to 7\,km \citep[see, for example,][]{kap08a}. The residuals mentioned above mostly contribute to the high value of chi-square. 

Similarly, if the residuals are not taken into account by adding more spectral complexity, atmosphere models do not provide good fits or plausible physical parameters for \jten. 
This is illustrated by fits (a) in Table~\ref{tab_resultspecNSA} and \ref{tab_resultspecNSMAX}, which consist of simply absorbed \textsf{\small nsa} and \textsf{\small nsmaxg} models respectively, with fixed element abundances (no \textsf{\small carbatm} model provided a fit with $\chi^2_\nu<10$). 
The fits lead to inconsistent column densities, sizes of the emission region, and distances; the quality is poor ($\chi^2_\nu=1.6-1.9$, for 166 to 169\,d.o.f.) and several parameters are unconstrained. The effective temperature of magnetised \textsf{\small nsa} models is usually above $10^6$\,K as they consist of fully-ionized hydrogen atmospheres. The only \textsf{\small nsmaxg} models with $\chi^2_\nu<2$ consist of cold, $T_{\rm eff}\sim(4-5)\times10^5$\,K, partially-ionized hydrogen atmospheres, under moderately strong magnetic fields of $B\sim(2-3)\times10^{13}$\,G. The large column density, $\nh\sim(5-7)\times10^{21}$\,cm$^{-2}$, is inconsistent with a best-fit distance of only $\sim60$\,pc. We verified that other element atmospheres (C, O, and Ne) in \textsf{\small nsmaxg} do not improve the spectral fitting. 

To test a multi-temperature spectral distribution, we next tried adding one extra thermal component to the absorbed blackbody and atmosphere models. With respect to (A), we obtain an improved fit quality (by $\Delta\chi^2_\nu\sim0.6$), and blackbody temperatures of $kT_1^\infty=28.6_{-2.1}^{+2.3}$\,eV and $kT_2^\infty=(117.8\pm2.2)$\,eV (B). However, the large column density of $\nh=(5.3\pm0.3)\times10^{21}$\,cm$^{-2}$ requires the model normalisation of the soft component to be large (the unabsorbed flux is nearly three orders of magnitude higher than that of a single-temperature blackbody). At 2.3\,kpc, the radiation radius of the soft component, $R_1^\infty\sim4,000$\,km, is inconsistent with that of a compact object. Even if the neutron star is in the foreground of the Carina complex, it should be at a maximum distance of 10\,pc from the Sun to constrain $R_1^\infty\lesssim20$\,km. Considering the large $\nh$ of the model, the low galactic latitude of the source, and a chance alignment with the Carina nebula, the solution is unlikely.

A second thermal component in the atmosphere models worsens the chi-square (b). Nonetheless, in spite of a rather poor fit ($\chi^2\sim1.7$ for 166 d.o.f.), one solution is more physical, the \textsf{\small nsa} model for $B=10^{12}$\,G, with $\nh\sim2.84_{-0.14}^{+0.20}\times10^{21}$\,cm$^{-2}$, $kT_1^\infty=34_{-13}^{+21}$\,eV and $kT_2^\infty=188.6_{-4}^{+1.9}$\,eV, at a best-fit distance of $\sim3.9$\,kpc. The best-fit radiation radii are, however, pegged at their boundary values of $R_1^\infty=20$\,km and $R_2^\infty=5$\,km. All other (b) models in Tables~\ref{tab_resultspecNSA} and \ref{tab_resultspecNSMAX} show large $\chi^2_\nu$, high $\nh$, and distances below $500$\,pc.

In (C)/(c), we found that the agreement between model and data is improved below $0.7$\,keV (typically by $\Delta\chi^2_\nu\sim0.3-0.5$), by allowing oxygen to be extrasolar, with $Z_{\rm O}\sim1.2-1.6$ in solar units, depending on the choice of continuum. In general, there is no significant change in the best-fit parameters (the exception is for the \textsf{\small nsa} model with for $B=10^{13}$\,G). Despite the improvement, all \textsf{\small nsmaxg} models (c) still show large $\nh$ and distances contradictorily below 100\,pc, regardless of the magnetic field strength.
The abundance of oxygen was the only one tightly constrained by the \textsf{\small vphabs} fits, which were found to be either insensitive to other elements (e.g.~C, Ne, Si, or Fe), or to produce large overabundances (e.g.~N, Mg, and Ca). In particular, elements with lines around $1.35$\,keV were verified to be arbitrarily overabundant in the model. A variable absorption model, combined with a double-temperature blackbody, has no effect on the elemental abundances, which are insensitive to the fit. This is likely due to the fact that the two blackbody components cross around an energy of $0.5$\,keV, where the \textsf{\small vphabs} model tries to accomodate for the extrasolar amount of oxygen.

Independently of the continuum -- single or double-temperature, blackbody or atmosphere, with fixed or overabundant elements -- we verified that the introduction of a Gaussian line in absorption around an energy of $1.35$\,keV significantly improves the quality of the spectral fitting. With respect to (C), the presence of the line improves the fit by $\Delta\chi^2_\nu\sim0.4$, resulting in a less absorbed model, with $\nh=2.17_{-0.20}^{+0.22}\times10^{21}$\,cm$^{-2}$ and $kT=(157\pm5)$\,eV (D).
The oxygen abundance is similar, $Z_{\rm O}=(1.64\pm0.12)$, and a best-fit line energy is centred at $\epsilon=1.289_{-0.017}^{+0.018}$\,keV, with equivalent width $EW\sim220$\,eV. 
Similar parameters and chi-square values are obtained when, alternatively, the abundances are fixed, and two lines in absorption are added to the blackbody continuum (E). In this case, the best-fit line energies are centred at $0.577_{-0.012}^{+0.013}$\,keV, and $1.361^{+0.015}_{-0.016}$\,keV and have equivalent widths of 90\,eV and 120\,eV. (To better constrain the model parameters, the Gaussian $\sigma$ of the two components were held fixed to 0.1\,keV and 0.2\,keV, respectively.)

As for (B), the presence of a line at $\epsilon=1.270_{-0.019}^{+0.018}$\,keV results in a lower flux normalisation and a column density consistent with that towards Carina, $\nh=3.4^{+0.4}_{-0.3}\times10^{21}$\,cm$^{-2}$ (F); the fit quality is improved by $\Delta\chi^2_\nu\sim0.3$ and the temperature of the two components are higher ($kT_1^\infty=36\pm4$\,eV and $kT_2^\infty=150\pm5$\,eV). Still, the neutron star should be at a maximum distance of 150\,pc to constrain $R_1^\infty\lesssim20$\,km.
Regarding the double-temperature atmosphere models, the presence of the absorption line at $\sim1.35$\,keV also results in solutions with more consistent physical parameters (d). In particular for \textsf{\small nsa} with $B=0$\,G and $B=10^{13}$\,G, the column density is $\nh=(2.6-2.8)\times10^{21}$\,cm$^{-2}$ and the distance is within $d\sim1.9$\,kpc to $3.3$\,kpc, i.e.~consistent with a neutron star in Carina. The best-fit solutions have $kT_1^\infty\sim30$\,eV and $kT_2^\infty=120-170$\,eV, with $R_1^\infty\sim20$\,km and $R_2^\infty\sim5$\,km.

Finally, we tested a combination similar to that in (E), i.e.~a single-temperature model with two lines, for the \textsf{\small nsa} and \textsf{\small nsmaxg} models. These are models (e) in Table~\ref{tab_resultspecNSA} and \ref{tab_resultspecNSMAX}, respectively, which are also the ones with the lowest-$\chi^2$ values in the analysis. 
Interestingly, all relevant parameters show best-fit values in a much narrower range. The energy of the lines are centred at $\epsilon_1\sim0.55$\,keV and $\epsilon_2\sim1.31-1.37$\,keV, and the equivalent widths are $EW_1\sim72-81$\,eV and $EW_2\sim200-300$\,eV. As for (E), the Gaussian $\sigma$ of the two components was fixed. Canonical $M_{\rm ns}$ and $R_{\rm ns}$ values typically provided the least chi-square fits. The size of the redshifted emitting region is within $8.5$\,km and $13$\,km -- overall consistent with the emission originating from most of the surface. The magnetic field intensity is not constrained by the atmosphere models, as the least-$\chi^2$ fits have $B=(0-3)\times10^{13}$\,G. In the lower plot of Figure~\ref{fig_bestBBfit}, we show the two pn spectra folded with the best-fit \textsf{\small nsa} model (e), for $B=10^{12}$\,G.

\begin{table}[t]
\caption{Best-fit parameters per observation and EPIC camera
\label{tab_bestfit}}
\centering
\begin{tabular}{l c r c c c}
\hline\hline
 & $\nh$\,\tablefootmark{$\dagger$} & \multicolumn{1}{c}{$T_{\rm eff}$} & \multicolumn{2}{c}{$\epsilon$} & $d$ \\ 
 & ($\times10^{21}$) & \multicolumn{1}{c}{($\times10^5$\,K)} & \multicolumn{2}{c}{(keV)} & (kpc) \\
\hline
(1) & $2.9_{-0.4}^{+0.5}$ & $9.3_{-0.9}^{+1.0}$ & \multicolumn{2}{c}{$0.569_{-0.025}^{+0.028},1.32(3)$} & $1.5_{-0.5}^{+0.6}$\\
(2) & $2.3_{-1.0}^{+1.2}$ & $9.1_{-2.1}^{+4}$   & \multicolumn{2}{c}{$0.516_{-0.021}^{+0.025},1.40(13)$} & $1.5_{-0.8}^{+1.5}$\\
(3) & $4.9_{-1.1}^{+1.5}$ & $8.6_{-1.7}^{+1.2}$ & \multicolumn{2}{c}{$0.80(5),1.24_{-0.07}^{+0.08}$}    & $0.9_{-0.3}^{+0.4}$\\
(4) & $2.2_{-0.5}^{+0.4}$ & $11.1_{-1.0}^{+1.6}$ & \multicolumn{2}{c}{$0.544_{-0.019}^{+0.022},1.295_{-0.025}^{+0.024}$} & $2.7_{-0.7}^{+0.6}$\\
(5) & $3.1_{-0.8}^{+0.9}$ & $9.4_{-1.6}^{+1.4}$ & \multicolumn{2}{c}{$0.59_{-0.03}^{+0.04},1.33(5)$} & $1.4_{-0.6}^{+0.9}$\\
(6) & $3.3_{-1.1}^{+1.3}$ & $9.1_{-1.4}^{+1.8}$ & \multicolumn{2}{c}{$0.52_{-0.04}^{+0.10},1.32(5)$} & $1.3_{-0.5}^{+0.7}$\\
\hline
\end{tabular}
\tablefoot{Errors are $1\sigma$ confidence levels. The fitted model is (e) for a \textsf{nsa} with $B=10^{12}$\,G (see~Table~\ref{tab_resultspecNSA}). The fitted spectra are (1) pn AO9, (2) MOS1 AO9, (3) MOS2 AO9, (4) pn AO11, (5) MOS1 AO11, (6) MOS2 AO11. The reduced chi-square values range between $0.6$ to $1.5$.
\tablefoottext{$\dagger$}{The column density is in cm$^{-2}$.}}
\end{table}
In Table~\ref{tab_bestfit} we summarise the best-fit parameters of this same model for the six individual EPIC spectra. The results are consistent overall. The most deviant parameters result from fitting the MOS2 spectrum of the AO9 observation. Indeed, this spectrum was not considered in the analysis of \citet{pir12}.

The inclusion of power-law tails extending towards higher energies has no effect on the parameters of the least-$\chi^2$ models. Considering again the best-fit \textsf{\small nsa} model (e), we found that non-thermal power-law components with typical photon indices of $\Gamma=1.7-2.1$ \citep[see, for example,][]{luc05} contribute at most 0.5\,\% ($3\sigma$ confidence level) to the luminosity of the source. 
\subsection{Timing analysis\label{sec_timinganalysis}}
To reassess the significance of the periodic signal detected at $P\sim18.6$\,ms, we restricted the timing analysis to the pn data and completely reanalysed the AO9 dataset consistent with the new \xmm\ observation. 
We considered events with photon patterns 12 or lower and tested different energy bands and radii of the source extraction region; the purpose was to unveil potential pulsations from the neutron star by minimising the background contribution. The spatial and energy selection cuts provide roughly $N_{\rm ph}\sim(4-6)\times10^3$ pn events in each observation. 

Each pn time series was Fourier-analysed in frequency domain to search for the presence of periodic signals. The $Z^2_n$ (Rayleigh) statistic \citep{buc83} was computed directly on the photon times-of-arrival; these were converted from the local satellite to the solar system barycentric frame using the SAS task \textsf{\small barycen} and the source coordinates in each pn observation (Table~\ref{tab_sourceMLparam}). We adopted a step in frequency of 2\,$\mu$Hz (or an oversampling factor of the expected width of the $Z^2_n$ peaks of nearly $10$), warranting that a periodic signal is not missed in our searches. The number of statistically independent trials in each $Z^2$ test is estimated as $\mathcal{N}=\Delta\nu t_{\rm exp}\sim8\times10^6$, with $\Delta\nu\sim87.7$\,Hz and $t_{\rm exp}\sim87$\,ks to $90$\,ks. In the absence of a periodic component, the mean and standard deviation of the $Z^2_n$ values, with $n=1$, are equal to 2, as the $Z^2_1$ statistic is distributed as a $\chi^2$ with two degrees of freedom. Therefore, the probability of obtaining a noise peak of a given power $Z^2$ in $\mathcal{N}$ independent trials is given by $\alpha=\mathcal{N}\exp(-Z^2/2)$; the confidence level of a detection can then be established as $C=(1-\alpha)\times100\%$.
\begin{figure}[t]
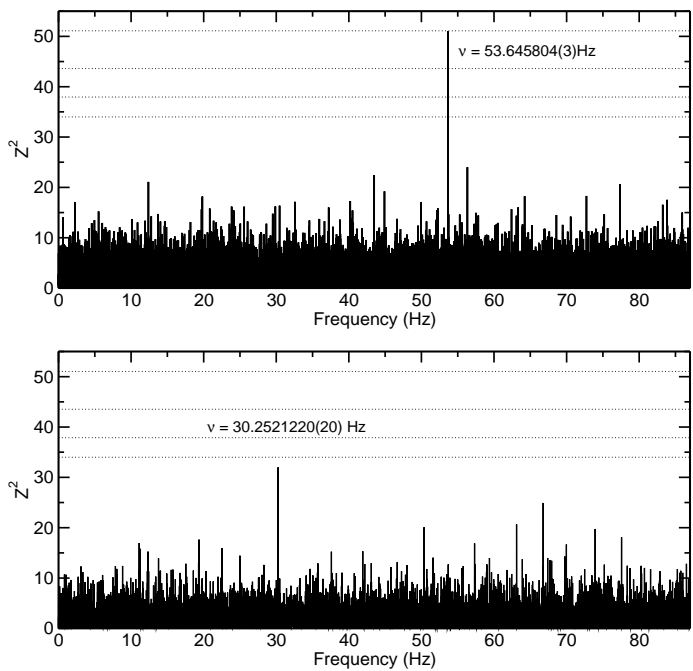

\begin{center}
\includegraphics*[width=0.49\textwidth]{fig3a.eps}\\\vspace{0.2cm}
\includegraphics*[width=0.49\textwidth]{fig3b.eps}
\end{center}
\caption{Results of $Z^2_1$ tests in the AO9 (top) and AO11 (bottom) pn time series. The frequency range is $\Delta\nu\sim87.7$\,Hz. Dashed horizontal lines show confidence levels of $1\sigma$ to $4\sigma$ ($68\%$ to $99.994\%$) for the detection of a periodic signal, taking the whole frequency range of the search into consideration (see text). The frequency of the highest peak resulting from each search is labelled. \textit{Top}: the energy band is $0.36-2.25$\,keV, the size of the extraction region is $18.6''$, and the number of photons is 5188. \textit{Bottom}: the energy band is $0.37-1.9$\,keV, the size of the extraction region is $19.7''$, and the number of photons is 5244. 
\label{fig_timing}}
\end{figure}

The analysis of the AO9 dataset yields only one statistically significant periodic signal, at $\nu_\ast=53.645804(3)$\,Hz ($P_\ast=18.6407869(12)$\,ms; see top plot of Fig.~\ref{fig_timing}). 
The power of the $Z^2_1$ statistic at $\nu_\ast$ is energy and signal-to-noise ($S/N$) dependent, being measured at best (or most significantly) when photons with energy below $\sim0.35$\,keV (where the noise of the pn camera in SW mode dominates) are excluded from the analysis, or when the source extraction region is around an optimal value between $17''$ and $22''$. These reflect the intervals where the $S/N$ ratio is optimal, indicating an origin on the neutron star itself. 

\begin{figure}[t]
\begin{center}
\includegraphics*[width=0.49\textwidth]{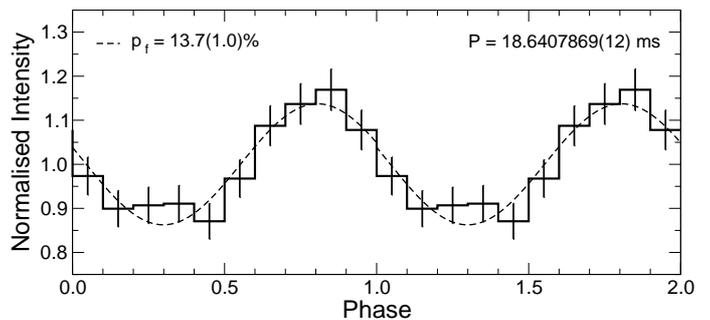}
\end{center}
\caption{AO9 pn light curve, folded at $P\sim18.6$\,ms (the epoch of phase zero is defined at $T_0 =$ MJD 5536.00884 days). The energy band is $0.36-2.25$\,keV and the size of the extraction region is $18.6''$. Two cycles are shown for clarity. The best-fit sinusoidal function is also shown (dashed line).
\label{fig_foldedLC}}
\end{figure}

The measured fluctuation of the power of the $Z^2_1$ statistic at $\nu_\ast$, $Z^2_1(\nu_\ast)\sim36-50$, is consistent with that expected from a sinusoidal modulation of amplitude $p_{\rm f}=(12.5\pm1.9)\%$ \citep[see, for example,][]{pav99}. Indeed, this is the amplitude derived from a sinusoidal fit of the light curve of the pn camera, folded at $P=P_\ast$ (Fig.~\ref{fig_foldedLC}); the pulsed fraction is also consistent with the non-parametric determination of the signal strength, $p_{\rm f}=(13.1\pm1.6)\%$, computed directly from the sample phases according to the bootstrap method proposed by \citet{swa96}. 

The inclusion of higher harmonics $n\ge2$ in the $Z^2_n$ test was found to be statistically insignificant.
Extensive searches showed no other significant signal (above the $1\sigma$ confidence level, or with $Z^2_1>34$) across the frequency range of the analysis, and safety checks did not suggest an unknown instrumental or background origin for the periodic signal at $\nu_\ast$ \cite[see][for details]{pir12}. We also verified that the periodic signal is always present at the same frequency, independently of the details on the processing of the raw event file (e.g.~included calibration files, SAS version, randomisation in energy within a PI channel, event filtering, or randomisation in time within the sampling detector time).

In the AO11 dataset, however, no obviously significant candidate frequency results from our $Z^2_n$ tests. Extensive tests positively exclude modulations with $p_{\rm f}>14\%$ ($Z^2_1>51$, $4\sigma$) in the $\nu=(0.01-87.73)$\,Hz frequency range of the search. However, at the brightness level of \jten, a modulation with $p_{\rm f}=11\%$ to $13\%$ may be missed in searches with too large a number of trials. As the background level in the second observation is higher and the exposure time slightly shorter, the pulsed signal at $P_\ast$\,ms may have been more severely affected (or smeared out) by noise. Alternatively, the AO9 signal is spurious and/or unrelated with the spin period of the neutron star (see discussion in Sect.~\ref{sec_sumresults}).

By considering $\Delta\nu=87.7$\,Hz, the only signal that shows a $Z^2_1$ power above the $1\sigma$ confidence level -- regardless of the energy band of the search -- is at $\nu_{\rm AO11}=30.252122(22)$\,Hz (see lower plot of Fig.~\ref{fig_timing}). 
The measured fluctuation in our tests is $Z^2_1=22-39$, consistent with a pulsed fraction of $p_{\rm f}=11.5(9)\%$. The overall signficance of the signal is low: on average, the probability of obtaining such a peak by chance is $\sim23\%$. Furthermore, the shift in frequency with respect to the first \xmm\ observation is too large ($\nu_\ast-\nu_{\rm AO11}\sim23.4$\,Hz) for the two signals to be associated. The implied spin-down rate would produce, within the usual magneto-rotational dipole-braking scenario in a vacuum, a magnetic field of $B_{\rm dip}=6.6\times10^{13}$\,G. In particular, a field with $B>5\times10^{12}$\,G would spin down the pulsar at a rate $>0.1$\,Hz\,yr$^{-1}$, making  the detection of a very short period of $\sim20$\,ms unlikely in the first place.

Regardless of the absence of an obvious candidate periodicity, the new \xmm\ observation can still be used in combination with the AO9 data in a coherent two-dimensional $Z^2_1(\nu,\dot{\nu})$ search, i.e.~allowing the $Z^2_n$ test to account for the neutron star spin-down. In this case, the ephemeris parameters $(\nu,\dot{\nu})$ that determine the phase $\phi_j$ of each photon time-of-arrival, 
\begin{displaymath}
\phi_j = \nu(t-t_0) + \dot{\nu}\frac{(t_j - t_0)^2}{2} \qquad j=1,\ldots,N_{\rm ph},
\end{displaymath}
\noindent are estimated as the values that give the highest power of the $Z^2_1$ statistic. 
Here, $t_0$ is the event time-of-arrival counted from an epoch of phase zero and $N_{\rm ph}=10\,355$ is the total number of photons in the search, when joining the two datasets for the optimal choices of energy band and extraction region.

We searched for significant $Z^2_1$ peaks in an interval of $\pm5\sigma$ around $\nu_\ast$, in steps of $5\times10^{-9}$\,Hz, and explored frequency derivatives from $-1\times10^{-18}$\,Hz\,s$^{-1}$ up to a maximum of $\dot{\nu}_{\rm max}=-1\times10^{-10}$\,Hz\,s$^{-1}$, in steps of $-2\times10^{-16}$\,Hz\,s$^{-1}$. The choice of the maximum frequency derivative is reasonable, as only three isolated pulsars -- the \object{Crab} and the energetic pulsars \object{PSR~J0537-6910} and \object{PSR~J0540-6919} in the Large Magellanic Cloud, out of about $2400$ known objects -- are known to slow down at a rate faster than $\dot{\nu}_{\rm max}$. The number of independent trials is given by $\mathcal{N}=(\nu_{\rm max} - \nu_{\rm min})|\dot{\nu}_{\rm max} - \dot{\nu}_{\rm min}|T_{\rm span}^{3}/2\sim4\times10^8$, where $T_{\rm span}=6.4\times10^7$\,s is the time span between the AO9 and AO11 observations. 

We found that the power of the $Z^2_1$ statistic increases significantly when folding the two datasets, even though the number of trials is boosted by a factor of $100$. The highest peak resulting from our search, with $Z^2_1\sim58$, occurs at $\nu=53.645805460(5)$\,Hz and $\dot{\nu}=-2.536200(15)\times10^{-11}$\,Hz\,s$^{-1}$ ($T_0 =$ MJD 55536.008840 days is the epoch of phase zero); the chance probability of the solution considering all number of independent trials is $0.01\%$ (confidence level of $3.885\sigma$). 
A total of 17 solutions, out of $\sim6\times10^9$, were found at or above the $3\sigma$ level (with $51.5\lesssim Z^2_1\lesssim58$); none was above the $4\sigma$ level. All high-$Z^2_1$ solutions have frequency derivatives within $-9.8\times10^{-11}\lesssim\dot{\nu}\lesssim-1.2\times10^{-11}$\,Hz\,s$^{-1}$. Within the scenario of magnetic braking in vacuum, the spin-down values imply a dipolar field of $B_{\rm dip}=(3-8)\times10^{11}$\,G for the source.

To test the result beyond its formal significance, we reproduced (in full resolution and over the same parameter space) the two-dimensional $Z^2_1(\nu,\dot{\nu})$ search, but folded AO11 events extracted from a background region over the original AO9 time series of the source. The background region was scaled to have roughly the same number of counts as collected for \jten, and two different energy bands were tested: the $0.36-2.2$\,keV of the original search, where the $S/N$ ratio is optimal, and the $3-10$\,keV energy band, which aims at excluding contamination from the neutron star. The goal is to verify empirically the power of the $Z^2_1$ statistic on a search with similar sensitivity, total duration, and time span of the original one, but conducted over the times of arrival of noise events that should not relate with the periodic signal at $\nu_\ast$.

We found that $Z^2_1$ peaks of similar power are obtained on these tests for intervals of frequency derivative that are roughly of the same order as those of the original search. In the $0.36-2.2$\,keV energy band, we found 14 solutions above the $3\sigma$ level (again out of $\sim6\times10^9$ grid points), with $Z^2_1\sim(51.5-61)$ and $-3\times10^{-11}\lesssim\dot{\nu}\lesssim-8\times10^{-12}$\,Hz\,s$^{-1}$. The highest $Z^2_1$ power occurs at $\dot{\nu}\sim-8\times10^{-12}$\,Hz\,s$^{-1}$ and implies $B_{\rm dip}\sim2.3\times10^{11}$\,G. For the $3-10$\,keV energy band, 28 solutions with frequency derivatives in the range $-7\times10^{-11}\lesssim\dot{\nu}\lesssim-1.2\times10^{-11}$\,Hz\,s$^{-1}$ were found with $Z^2_1\sim(51.5-60), $  and the highest power is for $\dot{\nu}\sim-7\times10^{-11}$\,Hz\,s$^{-1}$, which implies $B_{\rm dip}\sim7\times10^{11}$\,G. The tests show that the high $Z^2_1$ power is likely to be caused by statistical fluctuations in a search with a large number of trials and that the best $(\nu,\dot{\nu})$ solution derived for \jten\ is not significant.
\section{Discussion\label{sec_discussion}}
\subsection{Summary of results\label{sec_sumresults}}
At present, we cannot rule out the possibility that the periodic signal at $P_\ast\sim18.6$\,ms, which was only significantly detected in the AO9 observation ($p_{\rm f}=12.5(1.9)\%$, c.l.~$3\sigma$ to $4\sigma$, depending on the $S/N$ ratio of the search; see Sect.~\ref{sec_timinganalysis}) is unrelated to the true spin period of the neutron star.  
The second \xmm\ observation proved inconclusive at confirming the fast candidate spin, providing an upper limit that is close to the limiting sensitivity for detecting the AO9 modulation ($p_{\rm f}<14\%$ at $4\sigma$, for $P=0.0114-100$\,s and $0.35-2$\,keV). No other periodic signal was found to be statistically significant in extensive searches performed over the individual datasets (in a wide frequency range and for different tested energy bands). In particular, pulsations with $P>0.6$\,s (a range relevant for AXPs, the \msev\ INSs, and RRATs\footnote{As of 2015, over 1000 sources in the ATNF pulsar catalogue (of which 30 are X-ray pulsars and 58 are RRATs) are known to spin at periods longer than 0.6\,s \citep{man05}.}) are constrained down to $p_{\rm f}\sim10\%$ ($3\sigma$, $0.35-2$\,keV: \citealt{pir12}). Such smooth X-ray pulsations are usually more frequently observed in the \msev\ and CCOs, as AXPs typically show $p_{\rm f}\gtrsim15-30\%$.

The coherent combination of the two EPIC-pn time series, in a joint periodicity search around the candidate spin and accounting for a wide range of possible pulsar spin-down values, provides a $(P,\dot{P})$ solution with a probability of 1 in 10\,000 of it being spurious. Across the searched parameter space, the solutions above $3\sigma$ imply a dipolar magnetic field of $B_{\rm dip}=(3-8)\times10^{11}$\,G, under the assumption of magnetic braking in vacuum. However, additional tests suggest that the timing solutions are likely to be below their formal statistical significance.

In the absence of a clear period determination and good estimate of the magnetic field, the evolutionary state and nature of this neutron star remain as yet to be constrained. Unfortunately, at the flux level of the source and with current X-ray facilities, pulsations are difficult to detect below the limits provided by present data.

We were nonetheless able to study the spectral properties of \jten\ with unprecedented detail and statistics (Sect.~\ref{sec_spectralanalysis}). Interestingly, we found that good fits and meaningful physical parameters are only derived when the residuals at energies 0.55\,keV and 1.35\,keV are taken into account by the spectral modelling.
Independently of the exact choice of the thermal continuum, we found that the former residuals may result either from a local overabundance of oxygen in the Carina nebula (with $Z_{\rm O}=1.2-1.7$, relative to solar) or from the inhomogenous temperature distribution on the surface of the neutron star (the best-fit double-temperature models have $kT_1^\infty=30-36$\,eV and $kT_2^\infty=120-190$\,eV). Phenomenologically, the residuals at $0.55$\,keV can also be described by a Gaussian absorption line with $EW\sim70-90$\,eV. On the other hand, of all models tested in Sect.~\ref{sec_spectralanalysis}, the residuals at $1.35$\,keV are only satisfactorily accounted for by invoking a line in absorption, with $EW\sim120-400$\,eV (depending on the choice of the continuum). In this case, the fits lead to overall physically consistent (as well as much more tightly constrained) spectral parameters: $\nh=(2.5-3.3)\times10^{21}$\,cm$^{-2}$, $T^\infty=(5-8)\times10^5$\,K, $L_{\rm X}^\infty=(0.7-4)\times10^{32}$\,erg\,s$^{-1}$, and $d\sim1-3$\,kpc (considering the solutions with $\chi^2_\nu<1.4$).
\subsection{Comparison with other pulsars\label{sec_comparisonpulsars}}
Of all the INSs that clearly show thermal X-ray emission \citep[see, for example, the compilation of][]{vig13}, \jten\ displays spectral properties that are quite close to those of the \msev\ and CCOs -- namely, the soft energy distribution, constant X-ray properties on a long timescale ($\gtrsim15$\,yr), absence of significant magnetospheric emission, and evidence of absorption features. Deeper radio and $\gamma$-ray limits are, however, needed to exclude a middle-aged rotation-powered pulsar (RPP) or a mildly-recycled neutron star \citep[e.g.][]{bel10}.

The likely presence of \jten\ in the Carina open cluster excludes a neutron star much older than $\sim10^6$\,yr \citep[see discussion in][]{pir12}. 
In millisecond pulsars (MSPs), as well as in middle-aged and old RPPs, a considerable fraction of the X-ray emission arises from polar caps, which are heated by back-flowing charges accelerated in the pulsar magnetosphere. 
The spectral analysis shows that the X-ray emitting region in \jten\ is consistent with the surface of a canonical neutron star when realistic atmosphere models are adopted. This is in contrast to the $\sim0.1$\,km blackbody radii typically derived for MSPs \citep[e.g.][]{bog06,bog11}.

Middle-aged RPPs (e.g.~the Three Musketeers, with ages of a few $\sim10^5$\,yr and spin-down power of a few $\sim10^{34}$\,erg\,s$^{-1}$) are dominated by thermal emission and show much lower residual magnetospheric activity (at levels within $0.3$\,\% and $1.7$\,\% of the source luminosity; \citealt{luc05}). In \jten, no such weak hard tail (above $0.5\%$ of the source luminosity) is present, and we found no compelling evidence for a second (hotter) thermal component, as that typically arising from heated polar caps (see discussion in Sect.~\ref{sec_lines}).
For most RPPs, the efficiency at converting rotational energy into X-rays is generally above $\eta=10^{-5}$ \citep{kar12}. By considering the $3\sigma$ upper limit on the source non-thermal X-ray luminosity, $L_{\rm X}^{\rm pl}\lesssim2\times10^{30}$\,erg\,s$^{-1}$ (Sect.~\ref{sec_spectralanalysis}, for the best-fit \textsf{\small nsa} model (e) and $B=10^{12}$\,G), and assuming a maximum age of $10^6$\,yr for the neutron star, we can derive a lower limit on the spin-down luminosity of the source, $\dot{E}>2.1\times10^{36}$\,erg\,s$^{-1}$, if the neutron star is indeed a fast rotator. This implies an exceptionally low efficiency ratio of $\eta\lesssim10^{-6}$ for \jten, if the neutron star is a typical RPP. Alternatively, if the signal at $P_\ast$ is not related with the INS spin period, the possibility remains that \jten\ is a fainter and more distant \msev-like INS.
\subsection{Origin of absorption lines\label{sec_lines}}
The detection of spectral features in absorption in several (non-accreting) thermally emitting neutron stars -- including the \msev, CCOs, magnetars, and Calvera -- have been widely reported \citep[e.g.][]{san02,mor05,hab07,lau07,zan11,tie13}. 
The physical interpretation is not unique. The features can be due to atomic transitions in the magnetised atmosphere or condensed surface of the neutron star, or be cyclotron lines generated in a hot ionized layer near the surface. Importantly, the central energies of the features have been used to give estimates on the state of magnetisation of the neutron star. Recently, \citet{vig14} showed that, in some cases (as for the \msev\ \magze; \citealt{hab04b}), rather than an actual transition between energy levels, the spectral deviation may be induced simply by the inhomogeneous temperature distribution on the surface. 

In \jten, we verified that a double-temperature spectral model is only a plausible (physical) solution when we introduce an absorption feature at 1.35\,keV. Even in this case, the reduced chi-square values are still high (with respect to single-temperature models) and the radiation radii are uncertain. Possibly, the poor fit reflects the inadequate description of the anisotropic temperature distribution on the surface of a neutron star by the simple superposition of two atmosphere models.

If we assume that the feature at energy $\epsilon\sim1.35$\,keV is the fundamental electron cyclotron absorption, similar to the case of the CCO \object{1E~1207.4-5209} \citep{big03}, we can derive an estimate of the magnetic field on the surface for \jten. According to the relation $\epsilon_{\rm cyc}=1.16(B/10^{11}{\,\rm G})(1+z_{\rm g})^{-1}$, we obtain $B_{\rm cyc}\sim1.5\times10^{11}$\,G for the gravitational redshift of a canonical neutron star, $z_{\rm g}\sim0.3$. 
\subsection{Association with a runaway star in Carina}
A runaway massive star in the Carina nebula, \runaway\ (listed in the UCAC4 catalogue as 153-055048) was recently proposed to have been associated with the progenitor of \jten\ in a binary system, which was then disrupted by the supernova explosion that created the neutron star \citep{ngo13}. At a distance of $2.3$\,kpc, the $18'$ angular separation of the two objects on the sky implies a physical distance of $\sim12$\,pc. The total proper motion of the runaway star corresponds to a transverse speed of $v_{\rm t}=95\pm52$\,km\,s$^{-1}$ \citep{zac04}. Assuming typical masses of $1.4$\,M$_\odot$ and $9.6$\,M$_\odot$ (spectral type B1.5~V; \citealt{pec13}), respectively, for the neutron and runaway stars, the estimated flight time is of $t_{\rm kin}\sim(1.1-3)\times10^4$\,yr, which is roughly consistent with the absence of a supernova remnant.
Although uncertain, the implied velocity of \jten, $v_{\rm t}\sim650\pm360$\,km\,s$^{-1}$, if previously associated with \runaway, is also consistent with the distribution of transverse speeds observed in non-recycled radio pulsars \citep{hob05}. 
\subsection{Thermal evolution\label{sec_thermalevo}}
\begin{figure}[t]
\begin{center}
\includegraphics*[width=0.49\textwidth]{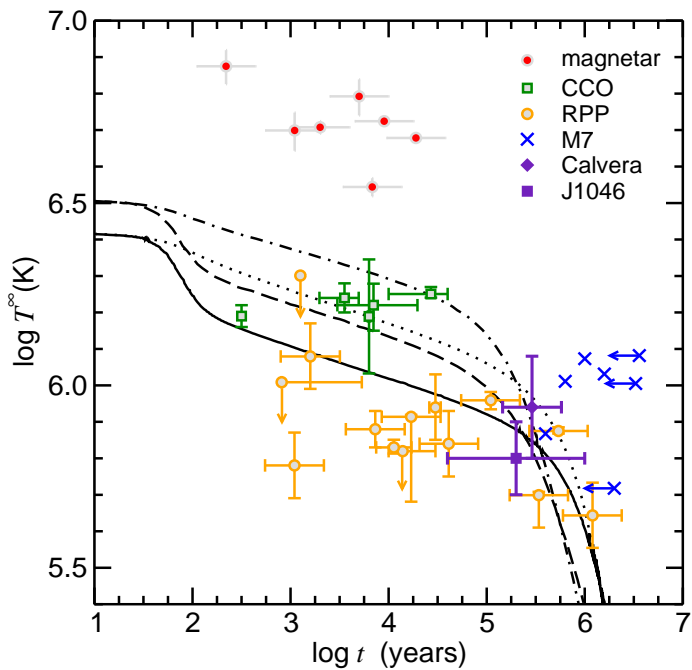}
\end{center}
\caption{Cooling-age diagram for different groups of isolated neutron stars (data points: see legend). Theoretical cooling curves for a canonical neutron star are shown as lines (see text). No effects of magnetic field decay are considered.\label{fig_ccurve}}
\end{figure}

Based on our best-fit spectral solutions, we can investigate the thermal state of the neutron star amidst the several INS populations for the first time. Figure~\ref{fig_ccurve} shows a representative sample of INSs, in a cooling-age diagram. 
The temperature $T^\infty$ is the apparent effective surface temperature, as derived from either atmosphere or blackbody models. 
Whenever possible, age estimates and intervals are derived from proper motion measurements (flight time) or associations with supernova remnants; otherwise the pulsar spin-down age, uncertain by a factor of two, is adopted. 
The data on RPPs (14 objects including the \object{Crab} and \object{Vela} pulsars, as well as the Three Musketeers and the old pulsar \object{PSR~J2043+2740}) are collected from \citet{yak08}. 
The five CCOs in the diagram are the neutron stars in the \object{Cas~A}, \object{Puppis~A}, \object{G296.5}, \object{Kes~79}, and \object{HESS~J1731-347} supernova remnants \citep[][and references therein]{bog14a,klo15}; in particular, their emission can be interpreted as radiation from the entire surface, using either hydrogen or carbon atmosphere models. 
The locus of magnetars (SGRs and AXPs) is represented by the seven objects investigated by \citet[][and references therein]{kam09}. Data points for the \msev\ are according to the compilations of \citet{kap11a} and \citet{vig13}. For these sources, characteristic ages are shown as upper limits on true ages in the absence of a kinematic estimate. 

Following our previous discussion, we adopt \jten's flight time as a lower limit for its age estimate; a conservative value of $10^6$\,yr is taken as an upper limit, based on an association with the Carina nebula. The temperature range is as reported in Section~\ref{sec_lines}. For Calvera, we adopt the temperature of the best-fit atmosphere (\textsf{\small nsa}) model \citep{zan11} and the INS characteristic age (again uncertain by a factor of two, \citealt{hal13}).

The set of theoretical cooling curves in Figure~\ref{fig_ccurve} correspond to four representative families, which are described in, for example, \citet{wei11,yak11,klo15}\footnote{As noted by \citet{klo15}, these curves are mostly aimed at explaining the residual heating observed in CCOs. Colder neutron stars, like some RPPs in Figure~\ref{fig_ccurve}, may be regarded as sufficiently massive stars with higher neutrino luminosity owing to the onset of the direct Urca process in the core.}. In a nutshell, the so-called standard neutrino candle, represented in Figure~\ref{fig_ccurve} by a solid line, shows the thermal evolution of a canonical neutron star (of mass $M=1.5$\,M$_\odot$ and radius $R=12$\,km, with a nucleon core), where cooling is governed by the modified Urca process, and an iron envelope is assumed (no enhanced cooling in the first 100 years). The dotted line shows the same star affected by strong proton superfluidity in the core, which highly suppresses the Urca reactions and results in a hotter neutron star at $t\gtrsim100$\,yr. The dashed and dot-dashed lines are the corresponding (non-superfluid and superfluid) models for a star with a full amount of accreted light elements (in particular, $\Delta M\sim10^{-8}$\,M$_\odot$ of carbon) in the blanketing envelope. The presence of light elements in the envelope of the neutron star increases the thermal conductivity \citep{pot97}; as a result, the light-element envelope regulates and slows down the cooling in the first $\sim10^4-10^5$\,yr, depending on the amount and composition of the accreted material. The effect of the magnetic field on cooling is not taken into account, which is a valid approximation for $B\lesssim10^{12}$\,G. 

The strong thermal emission (high temperature at a given age) observed in magnetars and the \msev, with respect to the standard cooling theory and other INS groups, cannot be explained without taking into account the effects of magnetic field decay as an additional source of heating of the neutron star crust. Magneto-thermal evolutionary models \citep{vig13} show that, for strongly magnetised INSs, field dissipation keeps the stellar crust hot for a longer time than expected from standard cooling \citep[see,][and references therein]{agu08}; moreover, strong fields at birth ($B>10^{14}$\,G) can significantly brake the neutron star spin to an asymptotic value in a relatively short timescale ($\sim10^5$\,yr). Interestingly, whereas for the bulk of the neutron star population such effects are negligible, the model implies an evolutionary link between the \msev\ and magnetars. 

On the other hand, consistent with the standard theory, CCOs are young neutron stars that show the slow cooling in the neutrino-cooling era typical of sources with light-element accreted envelopes. For these neutron stars, the turning point towards the steep drop in temperature of the photon-cooling stage takes place earlier than for neutron stars with an iron envelope.
As a result, old CCOs may be invisible to X-ray observations after $\sim1$\,Myr, where luminosities may have dropped below $\sim10^{31}$\,erg\,s$^{-1}$, and surface temperatures are colder than $\sim20$\,eV to 30\,eV. Nuclear burning in the envelope, and hence a change of composition, might nonetheless slow down the cooling in the last stages of the thermal life of the neutron star \citep{pag11}. 

Overall, the cooling status of Calvera appears consistent with a scenario where this neutron star evolved from a CCO (see Sect.~\ref{sec_intro}). However, Calvera's real surface temperature may be considerably softer than the adopted value in Figure~\ref{fig_ccurve}, $kT\sim98$\,eV, as the source shows a clear indication of a double-temperature energy distribution \citep{zan11}. In particular, the fit with a double thermal (blackbody or \textsf{\small nsa}) model constrains the apparent temperature of the cooler component within $T^\infty\sim(6-9)\times10^5$\,K, which is very similar to the interval estimated for \jten. 
Given its resemblance to Calvera, and if the fast spin and indication of a low magnetic field in \jten\ are to be confirmed, the INS in Carina may be another potential candidate for the elusive class of evolved anti-magnetars.
\section{Summary and conclusions\label{sec_summary}}
Following the intriguing detection of a short periodic signal in the neutron star \jtenfull\ \citep{pir12}, we targeted the source again with \xmm,\ aiming at better characterising its spectral energy distribution and timing properties. The new \xmm\ dataset unfortunately proved inconclusive in confirming the fast spin of the neutron star. The derived $4\sigma$ upper limit on the pulsed fraction, $p_{\rm f}<14\%$, shows that the observation is just at the limiting sensitivity to detect the modulation found previously.  

Without a clear determination of the spin period and a robust estimate of the magnetic field of the neutron star, its exact nature remains open to interpretation. The INS \jtenfull\ may be similar to Calvera, a neutron star for which the scenario of an evolved anti-magnetar, within the framework of the magnetic field burial by fallback of supernova debris, has been discussed. The overall spectral properties of \jtenfull\ and its likely presence in the Carina open cluster disfavour a recycled object or a standard evolutionary path, typical of that of a rotation-powered pulsar. However, better age estimates (e.g.~through kinematic studies), as well as deeper radio and $\gamma$-ray limits, are required to further constrain the evolutionary status of this neutron star. In particular, studies in the optical and near-infrared may offer the opportunity to assess properties which, for \jten, are not assessible by other means.

The prospect of finding more thermally emitting isolated neutron stars with the upcoming all-sky survey X-ray mission, \eROS\ \citep{pre14,mer12}, is of course of much interest. \eROS\ will scan the X-ray sky at unprecedented flux levels, making it possible to recognise orphaned CCOs and \msev-like neutron stars through their residual thermal emission, either among the known radio pulsar population or even if these elusive objects are intrinsically radio silent. 
\begin{acknowledgements}
We thank the anonymous referee for useful comments and suggestions that helped to improve the paper.
The work of A.M.P.~is supported by the Deutsche Forschungsgemeinschaft (PI~983/1-1). R.T.~is partially funded through an INAF PRIN grant. S.B.P.~was supported by the Russian Science Foundation, project 14-12-00146. The authors acknowledge the use of the ATNF Pulsar Catalogue \citep[][\texttt{http://www.atnf.csiro.au/research/pulsar/psrcat}]{man05}, the McGill Online Magnetar Catalog \citep[][\texttt{http://www.physics.mcgill.ca/\~{}pulsar/magnetar/main.html}]{ola14}, the RRATalog of discovered rotating radio transients (\texttt{http://astro.phys.wvu.edu/rratalog}), as well as the online catalogue of isolated neutron stars with significant thermal emission, described in \citet[][\texttt{http://www.neutronstarcooling.info}]{vig13}. 
\end{acknowledgements}
\bibliographystyle{aa}
\bibliography{ins}
\end{document}